\documentclass[journal]{IEEEtran}
\IEEEoverridecommandlockouts

\usepackage{cite}
\usepackage{amsmath,amssymb,amsfonts}
\usepackage{algorithmic}
\usepackage{graphicx}
\usepackage{textcomp}
\usepackage[dvipsnames]{xcolor}
\usepackage{spverbatim}
\usepackage{listings}
\usepackage{booktabs}
\usepackage{paralist}
\usepackage[symbols,nostyles,nomain]{glossaries}
\usepackage{glossary-super}

\makeglossaries	
\newglossaryentry{horizon}{%
	name={\ensuremath{N}},
	description={Prediction horizon in time steps},
	type=symbols
}

\newglossaryentry{softc}{%
	name={\ensuremath{\phi}},
	description={Soft constraint},
	type=symbols
}

\newglossaryentry{stagecost}{%
	name={\ensuremath{l}},
	description={Cost function},
	type=symbols
}

\newglossaryentry{terminalcost}{%
	name={\ensuremath{V}_f},
	description={Cost function},
	type=symbols
}

\newglossaryentry{model}{%
	name={\ensuremath{f}},
	description={Plant model},
	type=symbols
}

\newglossaryentry{uc}{%
	name={\ensuremath{h}},
	description={Control signal constraints function},
	type=symbols
}

\newglossaryentry{controlsignalvector}{%
	name={\ensuremath{\mathbf{u}}},
	description={Control signal vector},
	type=symbols
}

\newglossaryentry{xc}{%
	name={\ensuremath{g}},
	description={Plant state constraints function},
	type=symbols
}

\newglossaryentry{sigma}{%
	name={\ensuremath{\sigma}},
	description={Maximum control latency},
	type=symbols
}

\newglossaryentry{setpoint}{
	name={\ensuremath{x_s}},
	description={Set point},
	type=symbols
}

\newglossaryentry{tausc}{
	name={\ensuremath{\tau^{sc}}},
	description={Sensor-to-controller time},
	type=symbols
}

\newglossaryentry{tauca}{
	name={\ensuremath{\tau^{ca}}},
	description={Controller-to-actuator time},
	type=symbols
}

\newglossaryentry{tauc}{
	name={\ensuremath{\tau^c}},
	description={Computation time},
	type=symbols
}

\newglossaryentry{taua}{
	name={\ensuremath{\tau^a}},
	description={Admission time},
	type=symbols
}

\newglossaryentry{tau}{
	name={\ensuremath{\tau}},
	description={Request round-trip time},
	type=symbols
}

\newglossaryentry{valuefunc}{
	name={\ensuremath{V}},
	description={Value function},
	type=symbols
}

\newglossaryentry{dplantx}{
	name={\ensuremath{x}},
	description={Plant state},
	type=symbols
}

\newglossaryentry{dcontrolu}{
	name={\ensuremath{u}},
	description={Control signal},
	type=symbols
}

\newglossaryentry{contt}{
	name={\ensuremath{t}},
	description={Continous time},
	type=symbols
}

\newglossaryentry{dtime}{
	name={\ensuremath{k}},
	description={Discrete time in base frequency},
	type=symbols
}

\newglossaryentry{reqfreq}{
	name={\ensuremath{f_c}},
	description={Remote controller frequency},
	type=symbols
}

\newglossaryentry{tfl}{
	name={\ensuremath{\tau^{fl}}},
	description={Flight time, $\tau^{sc}+\tau^{ac}$},
	type=symbols
}

\newglossaryentry{Nt}{
	name={\ensuremath{N_t}},
	description={Prediction horizon in seconds},
	type=symbols
}

\newglossaryentry{gammax}{
	name={\ensuremath{\Gamma_x}},
	description={Plant state cost},
	type=symbols
}

\newglossaryentry{gammau}{
	name={\ensuremath{\Gamma_u}},
	description={Control signal cost},
	type=symbols
}

\newglossaryentry{gammas}{
	name={\ensuremath{\Gamma_{\phi}}},
	description={Soft constraint cost},
	type=symbols
}

\newglossaryentry{losssp}{
	name={\ensuremath{\rho_r}},
	description={Loss set-point},
	type=symbols
}

\newglossaryentry{err}{
	name={\ensuremath{e}},
	description={Controller input (error measure)},
	type=symbols
}

\newglossaryentry{ul}{
	name={\ensuremath{u_l}},
	description={Recovery control signal},
	type=symbols
}

\newglossaryentry{ustar}{
	name={\ensuremath{u^*}},
	description={MPC result vector},
	type=symbols
}

\newglossaryentry{A}{
	name={\ensuremath{A}},
	description={State transfer matrix},
	type=symbols
}

\newglossaryentry{B}{
	name={\ensuremath{B}},
	description={Control input transfer matrix},
	type=symbols
}

\newglossaryentry{reqloss}{
	name={\ensuremath{\rho}},
	description={Request loss},
	type=symbols
}

\newglossaryentry{qperiod}{
	name={\ensuremath{h_d}},
	description={Quantized sampling period},
	type=symbols
}

\newglossaryentry{emaalpha}{
	name={\ensuremath{\alpha}},
	description={Moving average weighting factor},
	type=symbols
}

\newglossaryentry{miss}{
	name={\ensuremath{l}},
	description={Miss ratio},
	type=symbols
}

\newglossaryentry{hc}{
	name={\ensuremath{h_c}},
	description={Frequency adaptation period},
	type=symbols
}

\newglossaryentry{pidk}{
	name={\ensuremath{K}},
	description={PID controller gain},
	type=symbols
}

\newglossaryentry{pidti}{
	name={\ensuremath{T_i}},
	description={PID integration time constant},
	type=symbols
}

\newglossaryentry{pidtd}{
	name={\ensuremath{T_d}},
	description={PID derivative time constant},
	type=symbols
}

\newglossaryentry{lqrk}{
	name={\ensuremath{L}},
	description={LQR gain matrix},
	type=symbols
}

\newglossaryentry{maxlag}{
	name={\ensuremath{\sigma}},
	description={Maximum control lag},
	type=symbols
}

\newglossaryentry{deadtime}{
	name={\ensuremath{D(r^x)}},
	description={Deadtime of request $r^x$},
	type=symbols
}

\newglossaryentry{lag}{
	name={\ensuremath{L(r^x_y)}},
	description={Lag of signal $y$ from request $r^x$},
	type=symbols
}

\newglossaryentry{sample}{
	name={\ensuremath{s}},
	description={A sample},
	type=symbols
}

\newglossaryentry{req}{
	name={\ensuremath{r}},
	description={A request},
	type=symbols
}

\newglossaryentry{stepsize}{
	name={\ensuremath{h_q}},
	description={Step size},
	type=symbols
}

\newglossaryentry{basefreq}{
	name={\ensuremath{f_b}},
	description={Base frequency},
	type=symbols
}

\newglossaryentry{mpcuvec}{
	name={\ensuremath{u^*}},
	description={MPC result vector},
	type=symbols
}

\newglossaryentry{response}{
	name={\ensuremath{\psi}},
	description={Response},
	type=symbols
}

\newglossaryentry{request}{
	name={\ensuremath{\phi}},
	description={Request},
	type=symbols
}	

\newcommand\fdeadtime[1]{\ensuremath{\mathcal{D}(#1)}}

\usepackage{soul}
\usepackage{amssymb}
\usepackage{amsmath}
\usepackage[utf8]{inputenc}
\usepackage{graphicx}
\usepackage{epstopdf}
\usepackage{nth}
\usepackage{float}
\usepackage{multicol}
\usepackage[printonlyused,withpage]{acronym} 
\usepackage{mathtools}
\usepackage{color}
\usepackage[inline]{enumitem}
\usepackage{footnote}
\usepackage{url}
\usepackage[capitalise]{cleveref}

\usepackage[normalem]{ulem}
\usepackage{siunitx}
\usepackage{etoolbox}
\usepackage{mathabx}
\usepackage{esvect}
\usepackage{booktabs}
\usepackage{subfigure}
\usepackage{csvsimple}
\usepackage{multirow}
\usepackage{xstring}
\usepackage{wrapfig}

\usepackage{tikz}
\usepackage{pgfplots}
\pgfplotsset{compat=1.15}
\pgfplotsset{compat/labels=pre 1.3}
\pgfplotsset{ylabsh/.style={every axis y label/.style={at={(0,0.5)}, xshift=#1, rotate=90}}}
\usepackage{varwidth}
\usetikzlibrary{plotmarks}
\usetikzlibrary{backgrounds}
\usetikzlibrary{fit}
\usetikzlibrary{graphs}
\usetikzlibrary{calc,positioning}
\usepackage{tikz-timing}
\usetikztiminglibrary[rising arrows]{clockarrows}
\usepackage{listofitems}
\usepgfplotslibrary{groupplots}
\usepgfplotslibrary{colorbrewer}
\usepgfplotslibrary[statistics]
\usepgfplotslibrary{external}
\usepgfplotslibrary{fillbetween}
\usetikzlibrary{patterns}
\usepackage{pgfplotstable}
\usepgflibrary{fpu}
\usetikzlibrary[fpu]

\usetikzlibrary{shapes.symbols}
\usetikzlibrary{fit}
\usetikzlibrary{topaths}

\usetikzlibrary{shadows.blur}
\definecolor{networkcolor}{rgb}{0.97,0.95,0.95}

\usetikzlibrary{chains,shapes.multipart}
\usetikzlibrary{shapes,calc,fit,arrows}
\usetikzlibrary{automata,positioning}
\usetikzlibrary{arrows.meta}

\usetikzlibrary{decorations.pathreplacing,decorations.markings,shapes.geometric,arrows,shapes,decorations,automata,backgrounds,petri}

\usetikzlibrary{graphs}

\makeatletter
\def\algbackskip{\hskip-\ALG@thistlm}
\makeatother

\makesavenoteenv{tabular}
\makesavenoteenv{table}

\DeclarePairedDelimiter{\ceil}{\lceil}{\rceil}

\newif\ifinsertannotations
\insertannotationsfalse 
\newif\ifinsertmargintodo
\insertmargintodofalse

\definecolor{diffaddcolor}{rgb}{1,0,0}
\definecolor{diffdelcolor}{rgb}{0.7,0.7,0.7}
\definecolor{diffshortencolor}{rgb}{0.2,0.5,0.2}

\renewenvironment{description}
{\list{}{\labelwidth=8pt \leftmargin=8pt
		}}
{\endlist}

\newcommand\tikzfiguresprefix{figures/}

\newcommand\insertfigure[2][generated]{%
	\includegraphics{gen/#2}
}

\newcommand\captionandlabel[2]{
	\caption{#1}\label{#2}
}

\tikzset{every picture/.append style={font=\footnotesize}}
\pgfplotsset{every axis legend/.append style={font=\footnotesize}}

\definecolor{calblue}{rgb}{0.337,0.478,0.51}
\definecolor{calgreen}{rgb}{0.471,0.733,0.259}
\definecolor{calorange}{rgb}{0.914,0.51,0.173}
\definecolor{calred}{rgb}{0.855,0.196,0.165}
\definecolor{wasppink}{RGB}{232,50,120}
\definecolor{waspgrey}{RGB}{66,80,82}
\definecolor{wasplightgrey}{RGB}{179,190,189}
\definecolor{waspblue}{RGB}{26,141,173}
\definecolor{waspbrightblue}{RGB}{146,219,239}
\definecolor{LUbronze}{rgb}{0.7,0.3,0}
\definecolor{LUblue}{rgb}{0,0,0.5}
\definecolor{LUpalegray}{rgb}{0.97,0.95,0.95}
\definecolor{LUalert}{rgb}{0,0,1}

\colorlet{color1}{calblue}
\colorlet{color2}{calgreen}
\colorlet{color3}{calorange}
\colorlet{color4}{calred}
\colorlet{color5}{waspgrey}
\colorlet{color6}{LUbronze!80}

\definecolor{calblue}{rgb}{0.337,0.478,0.51}
\definecolor{calgreen}{rgb}{0.471,0.733,0.259}
\definecolor{calorange}{rgb}{0.914,0.51,0.173}
\definecolor{calred}{rgb}{0.855,0.196,0.165}
\definecolor{wasppink}{RGB}{232,50,120}
\definecolor{waspgrey}{RGB}{66,80,82}
\definecolor{wasplightgrey}{RGB}{179,190,189}
\definecolor{waspblue}{RGB}{26,141,173}
\definecolor{waspbrightblue}{RGB}{146,219,239}
\definecolor{LUbronze}{rgb}{0.7,0.3,0}
\definecolor{LUblue}{rgb}{0,0,0.5}
\definecolor{LUpalegray}{rgb}{0.97,0.95,0.95}
\definecolor{LUalert}{rgb}{0,0,1}

\iftrue
\definecolor{clr-erdc00}{rgb}{0.645,0.745,0.626}
\definecolor{clr-erdc0}{rgb}{0.543,0.603,0.454}
\definecolor{clr-erdc1}{rgb}{0.453,0.443,0.311}
\definecolor{clr-erdc2}{rgb}{0.352,0.287,0.209}
\definecolor{clr-aws00}{rgb}{0.790,0.665,0.659}
\definecolor{clr-aws0}{rgb}{0.678,0.513,0.568}
\definecolor{clr-aws1}{rgb}{0.527,0.373,0.481}
\definecolor{clr-aws2}{rgb}{0.354,0.255,0.375}
\definecolor{clr-layer0}{rgb}{0.779,0.796,0.877}
\definecolor{clr-layer1}{rgb}{0.622,0.645,0.761}
\definecolor{clr-layer2}{rgb}{0.470,0.496,0.623}
\fi

\colorlet{mpccolor}{black!80!white}
\colorlet{ampccolor}{MidnightBlue!90!black}
\colorlet{oampccolor}{BrickRed!70!white}
\colorlet{rmpccolor}{OliveGreen!85!white}

\pgfplotscreateplotcyclelist{listone}{
	black!80!white,every mark/.append style={solid, fill=black},mark=*\\
	MidnightBlue!90!black,densely dotted,every mark/.append style={fill=MidnightBlue},mark=*\\	
	BrickRed!70!white,densely dashed, every mark/.append style={fill=MidnightBlue!20!white},mark=*\\
	OliveGreen!85!white,every mark/.append style={solid,fill=OliveGreen},mark=*\\	
	black!45!white, every mark/.append style={solid,fill=OliveGreen!20!white},mark=*\\	
	Lavender!80!black,densely dashed, every mark/.append style={solid, fill=black!20!white},mark=*\\			
}
\pgfplotscreateplotcyclelist{listonefill}{
	black,fill=black,every mark/.append style={solid, fill=black},mark=*\\
	MidnightBlue!90!black,densely dashed,fill=MidnightBlue!95!black,every mark/.append style={fill=MidnightBlue},mark=*\\
	BrickRed!70!white,densely dashed, fill=BrickRed!60!white, every mark/.append style={fill=MidnightBlue!20!white},mark=*\\	
	OliveGreen!85!white,fill=OliveGreen!85!white,every mark/.append style={solid,fill=OliveGreen},mark=*\\
	black!45!white, fill=black!45!white,every mark/.append style={solid,fill=OliveGreen!20!white},mark=*\\		
	Lavender!80!black,densely dashed, fill=Lavender!85!black, every mark/.append style={solid, fill=black!20!white},mark=*\\	
}

\pgfplotscreateplotcyclelist{listtwo}{
	black!80!white,every mark/.append style={solid, fill=black},mark=*\\
	MidnightBlue!90!black,every mark/.append style={fill=MidnightBlue},mark=*\\	
	OliveGreen!85!white,every mark/.append style={solid,fill=OliveGreen},mark=*\\
}

\pgfplotscreateplotcyclelist{listtwofill}{
	black,fill=black,every mark/.append style={solid, fill=black},mark=*\\
	MidnightBlue!90!black,densely dashed,fill=MidnightBlue!95!black,every mark/.append style={fill=MidnightBlue},mark=*\\
	BrickRed!70!white,densely dashed, fill=BrickRed!60!white, every mark/.append style={fill=MidnightBlue!20!white},mark=*\\	
	OliveGreen!85!white,fill=OliveGreen!85!white,every mark/.append style={solid,fill=OliveGreen},mark=*\\
	black!45!white, fill=black!45!white,every mark/.append style={solid,fill=OliveGreen!20!white},mark=*\\		
	Lavender!80!black,densely dashed, fill=Lavender!85!black, every mark/.append style={solid, fill=black!20!white},mark=*\\	
}

\pgfplotscreateplotcyclelist{listbwfill}{
	black,fill=black,every mark/.append style={solid, fill=black},mark=*\\
	black,fill=black,every mark/.append style={solid, fill=black},mark=*\\
	black,fill=black,every mark/.append style={solid, fill=black},mark=*\\
	black,fill=black,every mark/.append style={solid, fill=black},mark=*\\		
	black,fill=black,every mark/.append style={solid, fill=black},mark=*\\	
}

\newacro{5G}{Fifth Generation Wireless Specifications}
\newacro{AP}{Application Providers}
\newacro{API}{Application Program Interface}
\newacro{AR}{Augmented Reality}
\newacro{ARQ}{Automatic Repeat Query}
\newacro{AWS}{Amazon Web Services}
\newacro{BER}{Benefit Effort Ratio}
\newacro{BER}{Bit Error Rate}
\newacro{CAPEX}{Capital Expenditure}
\newacro{CDN}{Content Delivery Network}
\newacro{CLI}{Command Line Interface}
\newacro{CPS}{Cyber-Physical System}
\newacro{CPU}{Central Processing Unit}
\newacro{CRC}{Cyclic Redundancy Check}
\newacro{CSI}{Channel State Information}
\newacro{DB}{Database}
\newacro{DC}{Data Center}
\newacro{DoF}{Degrees of Freedom}
\newacro{DOF}{Degrees Of Freedom}
\newacro{EC2}{Elastic Compute Cloud}
\newacro{ECU}{electronic control unit}
\newacro{FaaS}{Function-as-a-Service}
\newacro{FEC}{Forward Error Correction}
\newacro{FPGA}{Field-Programmable Gate Array}
\newacro{GUI}{Graphical User Interface}
\newacro{HARQ}{Hybrid Automatic Repeat Query}
\newacro{HW}{Hardware}
\newacro{IaaS}{Infrastructure as a Service}
\newacro{i.i.d}{Independent and Identically Distributed random variables}
\newacro{IoE}{Internet of Everything}
\newacro{I/O}{Input/Output}
\newacro{IoT}{Internet of Things}
\newacro{IP}{Infrastructure Providers}
\newacro{KVS}{key-value store}
\newacro{LDPC}{Low-Density Parity-Check }
\newacro{LTE}{Long Term Evolution}
\newacro{LuMaMi}{Lund Massive MIMO}
\newacro{MAC}{Medium Access Control}
\newacro{MAN}{Metropolitan Area Network}
\newacro{MCN}{Heterogeneous Distributed Computing}
\newacro{MCN}{Mobile Cloud Network}
\newacro{MD}{Mobile Device}
\newacro{MD}{Mobile Devices}
\newacro{MEC}{Mobile Edge cloud}
\newacro{MIMO}{Multiple Inputs Multiple Outputs}
\newacro{mimo}[MIMO]{multiple-input-multiple-output}
\newacro{SISO}{Single Input Single Output}
\newacro{siso}[SISO]{single-input-single-output}
\newacro{MIP}{Mixed Integer Programming}
\newacro{mMTC}{massive Machine Type Communication}
\newacro{MN}{Mobile Network}
\newacro{MNO}{Mobile Network Operator}
\newacro{MNO}{Mobile Network Operators}
\newacro{MPC}{Model Predictive Controller}
\newacro{MQTT}{Message Queue Telemetry Transport}
\newacro{MR}{Maximum-Ration Combining}
\newacro{MT}{Mobile Terminal}
\newacro{MU-MIMO}{Multi-User MIMO}
\newacro{NFV}{Network Function Virtualisation}
\newacro{NoOps}{No Operations}
\newacro{OFDM}{Orthogonal Frequency-Division Multiplexing}
\newacro{OPEX}{Operational Expenditure}
\newacro{PaaS}{Platform as a Service}
\newacro{PDC}{Proximal Data Centers}
\newacro{PID}{Proportional Integral Derivative}
\newacro{PM}{Physical Machine}
\newacro{QoS}{Quality of Service}
\newacro{QPSK}{Quadrature Phase Shift Keying}
\newacro{RAN}{Radio Access Network}
\newacro{RAT}{Radio Access Technology}
\newacro{RBS}{Radio Base Station}
\newacro{RBS}{Radio Base Stations}
\newacro{RDC}{Remote Data Centers}
\newacro{RTD}{Round-Trip Delay time}
\newacro{RTT}{Round Trip Time}
\newacro{RTT}{Round-Trip Time}
\newacro{SaaS}{Software as a Service}
\newacro{SDK}{Software Development Kit}
\newacro{SDN}{Software Defined Networks}
\newacro{SDR}{Software Defined Radio}
\newacro{SLA}{Service Level Agreement}
\newacro{SLO}{Service Level Objective}
\newacro{SLO}{Service Level Objectives}
\newacro{SNR}{Signal-to-Interference-plus-Noise Ratio}
\newacro{SNS}{Simple Notification Service}
\newacro{SoS}{System of Systems}
\newacro{SP}{Service Providers}
\newacro{SQL}{Structured Query Language}
\newacro{SUMO}{Simulation of Urban MObility}
\newacro{SW}{Software}
\newacro{TLS}{Transport Layer Security}
\newacro{TraCI}{Traffic Control Interface}
\newacro{TSC}{Traffic Signal Control}
\newacro{TSP}{Transit Signal Priority}
\newacro{TTI}{Transmission Time Interval}
\newacro{UE}{User Equipment}
\newacro{UM}{User Mobility}
\newacro{URLLC}{Ultra-Reliable and Low-Latency Communication}
\newacro{UX}{User Experience}
\newacro{WAN}{Wide Area Network}
\newacro{WLAN}{Wireless Local Area Network}
\newacro{VM}{Virtual Machine}
\newacro{WSN}{Wireless Sensor Network}
\newacro{vSoftPLC}{virtual Software Programmable Logic Controller}
\newacro{PLC}{programmable logic controller}
\newacro{ZF}{Zero-Forcing}
\newacro{MS}{Mobile Station}
\newacro{NTP}{Network Time Protocol}
\newacro{PTP}{Precision Time Protocol}
\newacro{NGCC}{Next Generation Cloud Computing}
\newacro{ERDC}{Ericsson Research Data Cent´er}
\newacro{DNR}{Distributed-NodeRED}
\newacro{ADC}{Analog to Digital Converter}
\newacro{DAC}{Digital to Analog Converter}
\newacro{NoOps}{No-Operations}
\newacro{PaaS}{Platform-as-a-Service}
\newacro{WASP}{Wallenberg Autonoms Systems and Software Program}
\newacro{COTS}{Commercial off-the-shelf}
\newacro{COTC}[CotC]{Control over the Cloud}
\newacro{PoC}{Proof of concept}
\newacro{B'n'B}{Ball and beam}
\newacro{LQR}{Linear–Quadratic regulator}
\newacro{LQ}{Linear Quadratic}
\newacro{K8S}{Kubernetes}
\newacro{OSI}{Open Systems Interconnection}
\newacro{IIoT}{Industrial Internet-of-things}
\newacro{REST}{Representational state transfer}
\newacro{ICMP}{Internet Control Message Protocol}
\newacro{UDP}{User Datagram Protocol}
\newacro{TCP}{Transmission Control Protocol}
\newacro{CGI}{Common Gateway Interface}
\newacro{LAN}{Local Area Network}
\newacro{IT}{Information Technology}
\newacro{HTTP}{Hypertext Transfer Protocol}
\newacro{HA}{High availability}
\newacro{RAV}{Relative Accumulated Violations}
\newacro{RAE}{Relative Accumulated Error}
\newacro{RMCV}{Relative Maximum Constraint Violation}
\newacro{EDC}{Edge Data Center}
\newacro{TSN}{Time Sensitive Networks}
\newacro{ICT}{Information and Communications Technology}
\newacro{CCS}{Cloud Control System}
\newacro{NCS}{Networked Control System}
\newacro{MLE}{Maximum Likelihood Estimation}
\newacro{R-CCS}{Resilient Cloud Control System}
\newacro{AGV}{automated guided vehicles}
\newacro{CLRE}{closed loop response error}
\newcommand{\stoawsregion}{eu-north-1}
\newcommand{\stoawsloc}{Stockholm, Sweden}
\newcommand{\fraawsregion}{eu-central-1}
\newcommand{\fraawsloc}{Frankfurt, Germany}

\newcommand{\plant}{plant}

\graphicspath{{figures/},\tikzfiguresprefix}

\IfFileExists{localheader.tex}{\input{localheader}}

\providecommand{\ruggedtodo}[2][]{}

\newcommand{\ke}[2][]{%
\IfStrEq{#1}{fixed}{{\color{green!80!black}\small{\textbf{Karl-Erik:} #2}}}{{\color{magenta!80!black}\small{\textbf{Karl-Erik:} #2}}}%
}

\usepackage{tcolorbox}
\colorlet{clr-per}{OliveGreen}
\definecolor{clr-william}{rgb}{0.452,0.201,0.401}
\colorlet{clr-maria}{Dandelion}
\colorlet{clr-ke}{BrickRed}
\colorlet{clr-review}{Bittersweet!90!white}
\colorlet{clr-rewrite}{BrickRed}
\colorlet{clr-done}{OliveGreen}

\ifinsertannotations
	\newcommand\reviewper[1]{\tcbox[on line,boxsep=0pt,boxrule=0pt,left=3pt,right=3pt,top=2pt,bottom=2pt,colback=clr-per,coltext=white]{Per:} {\it\color{clr-per}#1}}
	\newcommand\reviewwilliam[1]{\tcbox[on line,boxsep=0pt,boxrule=0pt,left=3pt,right=3pt,top=2pt,bottom=2pt,colback=clr-william,coltext=white]{William:} {\it\color{clr-william}#1}}
	\newcommand\reviewmaria[1]{\tcbox[on line,boxsep=0pt,boxrule=0pt,left=3pt,right=3pt,top=2pt,bottom=2pt,colback=clr-maria,coltext=white]{Maria:} {\it\color{clr-maria}#1}}
	\newcommand\reviewke[1]{\tcbox[on line,boxsep=0pt,boxrule=0pt,left=3pt,right=3pt,top=2pt,bottom=2pt,colback=clr-ke,coltext=white]{KE:} {\it\color{clr-ke}#1}}
	\newcommand\reviewtodo[2][Review:]{\tcbox[on line,boxsep=0pt,boxrule=0pt,left=3pt,right=3pt,top=2pt,bottom=2pt,colback=clr-review,coltext=white]{#1} {\it\color{clr-review}#2}}
	\newcommand\reviewrewrite[2][Rewrite:]{\tcbox[on line,boxsep=0pt,boxrule=0pt,left=3pt,right=3pt,top=2pt,bottom=2pt,colback=clr-rewrite,coltext=white]{#1} {\it\color{clr-rewrite}#2}}
	\newcommand\reviewdone[1][Done]{\tcbox[on line,boxsep=0pt,boxrule=0pt,left=3pt,right=3pt,top=2pt,bottom=2pt,colback=clr-done,coltext=white]{#1}{\it\color{clr-done}}}

	\newcommand{\per}[1]{\reviewper{#1}}
	\newcommand{\william}[1]{\reviewwilliam{#1}}
	\newcommand{\maria}[1]{\reviewmaria{#1}}	
	\providecommand{\ke}[1]{\reviewke{#1}}
	\newcommand{\rewrite}[1]{\reviewrewrite{#1}}
	\newcommand{\review}[1]{\reviewtodo{#1}}
\else
	\newcommand{\per}[1]{}
	\newcommand{\william}[1]{}
	\newcommand{\maria}[1]{}	
	\providecommand{\ke}[1]{}
	\newcommand{\rewrite}[1]{}
	\newcommand{\review}[1]{}
	
	\newcommand\reviewper[1]{}
	\newcommand\reviewwilliam[1]{}
	\newcommand\reviewmaria[1]{}
	\newcommand\reviewke[1]{}
	\newcommand\reviewtodo[2][Review:]{}
	\newcommand\reviewrewrite[2][Rewrite:]{}
	\newcommand\reviewdone[1][Done]{}
	
\fi

\RequirePackage[normalem]{ulem} 
\RequirePackage{color}\definecolor{RED}{rgb}{1,0,0}\definecolor{BLUE}{rgb}{0,0,1} 
\ifinsertannotations
\providecommand{\DIFadd}[1]{{\protect\color{diffaddcolor}#1}}
\providecommand{\DIFdel}[1]{{\protect\color{diffdelcolor}\sout{#1}}} 
\providecommand{\DIFshorten}[1]{{\protect\color{diffshortencolor}\sout{#1}}} 
\else
\providecommand{\DIFadd}[1]{#1}
\providecommand{\DIFdel}[1]{}
\providecommand{\DIFshorten}[1]{} 
\fi

\providecommand\replace[2]{\DIFdel{#1}\DIFadd{#2}}

\ifinsertannotations
\makeatletter
\def\uwave{\bgroup \markoverwith{\lower3.5\p@\hbox{\sixly \textcolor{red}{\char58}}}\ULon}
\makeatother

\else

\fi

\ifinsertmargintodo
\usepackage[color=black!20!white,backgroundcolor=black!5!white,bordercolor=black!20!white]{todonotes} 
\else
\fi

\newif\ifshowremoved
\showremovedfalse
\newcommand\markstartremoved{
	\begingroup
	\color{red}	
	\par\noindent\rule{\columnwidth}{0.4pt}
	REMOVED
}
\newcommand\markendremoved{\par\noindent\rule{\columnwidth}{0.4pt}\endgroup}

\def\BibTeX{{\rm B\kern-.05em{\sc i\kern-.025em b}\kern-.08em
    T\kern-.1667em\lower.7ex\hbox{E}\kern-.125emX}}

\begin{document}
	
\acused{PID}
\acused{LQR}

\title{ \huge
	Resilient Cloud Control System: Realizing resilient cloud-based optimal control for cyber-physical systems
  \thanks{
    This work has been partially funded by the Wallenberg AI, Autonomous Systems and Software Program (WASP), the ELLIIT strategic research area on IT and mobile communications, Sweden's Innovation Agency (VINNOVA) under the 5G-PERFECTA Celtic Netxt project, the Swedish Foundation for Strategic Research under the SEC4FACTORY project.
    }
  }

\author{
    \IEEEauthorblockN{
      William Tärneberg\IEEEauthorrefmark{2},
      Per Skarin\IEEEauthorrefmark{1}\IEEEauthorrefmark{3},
      Karl-Erik Årzén\IEEEauthorrefmark{1}, and
      Maria Kihl\IEEEauthorrefmark{2}
      } \\
    \IEEEauthorblockA{Departments of \IEEEauthorrefmark{1}Automatic Control \& \IEEEauthorrefmark{2} Electrical and Information Technology, at Lund University, Sweden}
    \IEEEauthorblockA{\IEEEauthorrefmark{3}Ericsson Research, Lund, Sweden}    

}

\maketitle

\begin{abstract}
The transformation to smart factories and the automation of mobile robotics is partly driven by a growing availability of ubiquitous cloud technologies. In cyber-physical systems, such as control systems, critical parts can be migrated to a cloud for offloading, enabling collaborative processes, improved performance, and life-cycle management. Despite the performance uncertainty in a cloud and the intermediate networks, presently, even cloud native function services are being investigated for supporting critical applications that are sensitive to time-varying execution and communication delays. In this paper, we introduce, implement, and empirically evaluate an architecture that successfully allows predictive controllers to take advantage of cloud native technology. Our solution relies on continuously adapting the control system to the present Quality of Service of the cloud and the intermediate network. As our results show, this allows a control system to survive interruptions, noisy neighbors, and time-variant resource availability. Without the proposed solution, the control system will fail due to resource constraints and insufficient response times. Further, we also show a system that can seamlessly switch between clouds and that multiple controllers using shared resources consequentially self-adapt so that no controller fails its objective.

\end{abstract}


%
%

\section{Introduction} \label{sec:introduction}

	Process control in smart factories, \ac{AGV}, gaming, augmented reality, and smart grids are examples of areas in which time-critical systems are aiming to co-exist with and utilize cloud services. 
	There, latency sensitive parts of applications are migrated to the cloud to gain access to data, economies of scale, offloading, collaboration, and efficient software development practices.
	But the cloud is a problematic environment for mission-critical and time-sensitive applications because of reliability, availability, and the time-varying execution and communication delays that are inherent properties of cloud deployments.

    To enable control systems to successfully execute on cloud platforms, in addition to making networks, software platforms, and services support real-time requirements, controllers should be made adaptive to the prevailing state of the cloud infrastructure, employing the principle of quality elasticity \cite{larsson2019quality}. Therein lies the focus of this paper.
    To be more precise, \acp{CCS} must be able to handle various delay distributions, uncertain processing times, different types of deployments, and make positive use of resource scaling.
    These relevant and unique properties of \acp{CCS} have received limited attention from the control community.
    
    While there are few works on critical systems that specifically consider a reactive approach to resource scaling of clouds, there are several works on control over shared networks that implement on-line adaptation for sharing and resource preservation. 
    Two examples are~\cite{ploplys2004closed} and \cite{xia2007flexible}.
    However, the literature has several shortcomings in relation to cloud, such as
    \begin{inparaenum} 
        \item assuming a single delay distribution,
        \item assuming limited range in the variability of delays, 
        \item solving for a specific control problem,
        \item considering only the network as a source of error, and
        \item a focus on simple controllers, such as PIDs, that are not intrinsically useful in the cloud context.
    \end{inparaenum}
    \reviewwilliam{We do not circle back to this above research-gap in the paper.}
    
    In this paper, we address the above shortcomings and construct a fully functional resilient and adaptive \ac{CCS}, the \ac{R-CCS}. 
    \ac{R-CCS} takes a general approach to uncertainty and acknowledges that a \ac{CCS} cannot be measured as mandated by the literature. Further, \ac{R-CCS} is a composition of multiple remedies that makes \ac{R-CCS} adaptable to many control problems. In this paper, we apply \ac{R-CCS} using Model Predictive Control (MPC)\acused{MPC}, a key ingredient in modern automatic control\replace{ that is more demanding and susceptible to the uncertainties in cloud than previously seen in the literature}.
    We proceed, through experimentation, to demonstrate that the controller can effectively handle network delays and that the predictive controller can mitigate its own extensive and variable computation time.
    Further, we show that a set of controllers can run on a shared infrastructure with the duty cycle requirements of a factory automation system (i.e. \SI{10}{Hz} to \SI{1}{kHz}~\cite{khan2017analysis}) if they implement the proposed adaptive structure, referred to as \ac{R-CCS}. 

    The primary contribution of the paper is the proposal of an \ac{R-CCS} to incorporate the benefits of using cloud platforms and cloud native software into the domain of critical control systems. 
    We deploy an offloaded \ac{MPC} controller and expose it to three challenges defined in \cref{sec:challenges}.
    Through simulations and cloud deployment we show that the \ac{R-CCS} can mitigate all three challenges.
    Specific contributions to achieve this goal are:
    \begin{enumerate}
        \item design and evaluation of a variable rate \ac{MPC},
        \item a concrete design for a cloud native \ac{R-CCS},
    	\item simulations based on data from previous experiments with cloud native \ac{MPC} deployment,
        \item a general and multi-faceted evaluation of the \ac{R-CCS}'s resiliency and its ability to adapt to circumstances.
    \end{enumerate}

\section{Method} \label{sec:preliminaries}
In this section, we detail the model and challenges of a \ac{CCS}.
We then present our \ac{R-CCS} based on four remedies, followed by a set of simulations that motivate the use of the remedies.

\subsection{Problem}

\begin{wrapfigure}{R}{0.6\columnwidth}
            \centering
            \vspace{-.5em}
    			\begin{tikzpicture}[>=stealth,line width=0.75pt,font=\footnotesize]
    				\node[draw,fill=white] (controller) {Controller};
    				\node[above=0.3 of controller] (obj) {Objective};
    				\node[draw,fill=white,below=0.3 of controller] (plant) {Process};
    				\node[draw,fill=white,right=0.5 of plant] (act) {Actuator};
    				\node[draw,fill=white,left=0.5 of plant] (sens) {Sensors};
    				\draw[->] (obj) -- (controller);	-
    				\draw[->] (sens) |- (controller);
    				\draw[->] (sens) -- node[above,pos=1] {Monitor} ($(sens|-obj.south)+(0,0.1)$);
    				\draw[->] (controller) -| (act);
    				\draw[->] (act) -- (plant);
    				\draw[->] (plant) -- (sens);
    				
    				\begin{scope}[on background layer]
    					\coordinate (ract) at ($(act.south east)+(0.1,-0.1)$);
    					\coordinate (lact) at ($(act.south west)+(-0.2,-0.1)$);
    					\coordinate (rsens) at ($(sens.south east)+(0.2,-0.1)$);
    					\coordinate (lsens) at ($(sens.north west)+(-0.1,0.15)$);
    					
    					\coordinate (start) at ($(sens.west|-controller.north)+(-0.1,0.2)$);
    					\draw[gray,fill=gray!5!white] (start) -- (start-|ract) -- (ract) -- 
    						(lact) -- (lsens-|lact) -- (rsens|-lsens) -- (rsens) --
    						(start|-rsens) -- (start);
    					\node[anchor=north east] at (start-|ract) {\textit{PLC/ECU}};
    						
    %
    				\end{scope}
    			\end{tikzpicture}
    		\vspace{-0.5em}
    			\caption{The basic control loop}
    			\label{fig:cloop}
                \vspace{-0.8em}
	\end{wrapfigure}
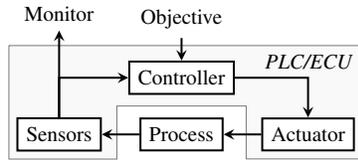

    Elementary, a \ac{CCS} \cite{skarin2020cloudperformance} is a system where parts of the \emph{controller} logic is deployed on or offloaded to a cloud platform.
    The basic control loop is illustrated in \cref{fig:cloop}.
    \cref{fig:cloop} shows a process that is being controlled using a \ac{PLC} or \ac{ECU}. 
    Through its actuators, the controller is in charge of manipulating the state of the process, also referred to as the plant, to achieve some objective.
    Many controllers are in themselves primitive and light-weight in terms of computations, such as the \ac{PID} and \ac{LQR}, but others, such as, numerical optimizations and machine learning tasks, can be resource intensive.
    In network control, light-weight controllers are often considered, and communication delay is of primary interest while execution is assumed fast (and straight forward to prioritize).
    \emph{Network control theory} studies the limitations that apply to the communication channel and controller in order for the overall closed loop system to meet the objectives and remain reliable.
    In networked control, system load is often defined in terms of transferred data, i.e., the traffic load on the network. 
    In a \ac{CCS}, the execution time of the controller in the cloud and the admission of samples to the cloud also need to be considered.

    Our objective is to move the controller part of \cref{fig:cloop} into the cloud as a micro-service in a \ac{FaaS} deployment. This allows us to take advantage of the compute capacity and scalability of the cloud.
    We therefore deploy an \emph{offloading predictive controller} - a controller that uses numerical optimization to find optimal control actions and which executes on a remote host, such as in the cloud.

    Management logic relating to the system's objectives is implemented at the plant.
    It can be implemented in the \ac{PLC}/\ac{ECU} in \cref{fig:cloop} but this can also be replaced by a more primitive \ac{IoT} device.
    The \emph{state} is sent over a network to the remote controller and expects a \emph{control signal} in return.     
    The remote controller is implemented as a web-service, accessed over \acs{HTTP}, and packaged as a micro service in a container or as a \ac{FaaS} function.
    The implementation of the controller is identical in these two scenarios.
 	The deployment of the controller into an public cloud subjects the communication between the plant and the remote controller to stochastic delays and the controller to variations in execution time.  
    The scale and impact of such delay, at various levels in the cloud stacks, was explored in \cite{skarin2020cloudperformance}.  
    Upon deployment, no prior knowledge of the infrastructure is assumed, and it is expected that the properties of networks and execution platforms are time-variant.
    


    \emph{Resilience} and \emph{adaptability} are key design goals for a \ac{CCS}. 
    A resilient \ac{CCS} shall be able to provide:
    \begin{enumerate*}
        \item good performance in environments that provide real-time support, and 
        \item reliable execution in environments without real-time support, and 
        \item ability to cope with the risk of loss of connectivity and other service failures. 
    \end{enumerate*}
    This is true whether using public or private clouds and is a consequence of the flexibility cloud platforms offer its operators and users.
    An ideal \ac{CCS} is not explicitly designed for a certain state or observed performance level of an infrastructure but instead adapts to the prevailing circumstances of any deployment.
    In a sense, a \ac{CCS} shall have the ability to be \emph{quality elastic}, meaning that it can adjust its performance to match available resources.

     \subsubsection{Cloud Control Challenges}\label{sec:challenges}
	 Inherited from networked control and real-time systems, a \acl{R-CCS} has to address synchronization between the cloud and its clients. Further, below we identify four primary challenges.
	 
	 \begin{description}
	 \item[Resilience]{
	 A \ac{R-CCS} shall be tolerant to non-trivial delay distributions and connectivity issues. We refer to the latter as partition tolerance, something that is generally not applicable to network control systems.
	 Partition tolerance necessarily requires that the system remain in a recoverable state, not requiring manual intervention, even if the remote extension becomes unresponsive indefinitely.}
	 
	 \item[Performance]{
	 The \ac{R-CCS} should not rely on fault tolerance mechanisms that systematically penalize or degrade general performance when the system is not subject to loss.
	 Under nominal circumstances, the control system shall achieve good, nearly optimal, performance.}
	 
	\item[Costs]{
	Events such as lost connections, service errors, unsolvable optimization problems, or long delays, are assumed to occur frequently and are handled as failed requests. They constitute resource waste that incur a monetary cost and can have adverse system effects. 
	The system should work to reduce resource waste, while balancing its performance.}

    \item[\ac{COTS}]{
    The implementation and deployment shall be portable across cloud platforms and rely on the boiler-plate code and function provided by the cloud platforms. 
    To that end, the implementation has to rely on \ac{COTS} and follow cloud-native development practices.
    }
 	\end{description}
	    
   
    \subsubsection{Offloading Numerical Optimizations} \label{sec:controller}
        This paper focuses on offloading control to the cloud.
        The research platform achieves a basic control task but implements a useful practice that extends directly to more complex problems through the \ac{MPC}. 
        An \ac{MPC} generates its control signal by repeatedly solving an optimization problem.
        We write the basic \ac{MPC} problem as

\begin{equation}
	\begin{aligned}
		\text{minimize}\quad & J(z), \text{ from } t_0 \text{ to } t_N\\
		\text{subject to}\quad &f(z) = 0,\ g(z) \le 0
	\end{aligned}\label{eq:mpc}
\end{equation}
where $J$ is a cost function defining the control objectives, $f$ provides a plant model,  and $g$ defines system limits.
Time $t_0$ to $t_N$ represent a time frame over which the controller predicts its input and the state of the plant. 
$N$, representing a number of time steps, is referred to as the controller horizon.
The optimization variable $z$ contains a series of predicted system states $\begin{bmatrix}x(0) & \dots & x(N)\end{bmatrix}$ and control actions $\begin{bmatrix}u(0) & \dots & u(N-1)\end{bmatrix}$, where $x$ and $u$ are vectors of arbitrary length.
With every sample read from the sensors (at time $k$), the controller uses \cref{eq:mpc} to find an optimal sequence of control actions $u_k$.
The nominal \ac{MPC} applies the first action in this sequence  $u_k(0)$ and repeats the procedure.

The time it takes to find $u$ depends on the detailed implementation of \cref{eq:mpc}, the method used to solve it, and the size of the controller horizon. 
An example for which efficient solvers exists is the quadratic program
\begin{equation}
	\begin{aligned}
		\text{minimize}\quad &z^T H z + h z\\
		\text{subject to}\quad &T z = t,\ G z \le g,
	\end{aligned}\label{eq:mpc2}
\end{equation}
where $H$, $h$, $T$, $t$, $G$, and $g$ are matrices and vectors of sizes determined by $N$ and the dimensions of $x$ and $u$.
Nonetheless, even such an implementation can be relatively computationally intensive and exhibit variable execution times due to an unknown and changing number of iterations required to solve the optimization problem.
The used controller implements the quadratic program and adds soft constraints to form
\begin{equation}
	g(x, \phi) \le 0,\quad z =  \begin{bmatrix}x^T & u^T & \phi^T\end{bmatrix}^T.\\\\
\end{equation}
Soft constraints allow constraint violation with a large penalty in the cost function.
This is a common extension to \ac{MPC} and refer the reader to the extensive literature on the subject.
 
Usually, an \ac{MPC} is implemented and tested for a tailored execution environment, the size and complexity of \labelcref{eq:mpc} does not change during run-time, and the implementation is expected to always provide a result within a fixed time. \reviewwilliam{@PER!: Are we clear about this in the rest of the paper: that we kinda break the expetations of an MPC when we put it in the cloud?` Perhaps mention earlier.}
However, since the controller is executing an explicit optimization, there is potential for flexible variation of these properties over time.
\replace{In the traditional setup, this would}{This can} be limited to properties that do not change the basic complexity, such as modifying the parameters in the model, the cost function, or the system limits, as respectively defined by $(T,t)$, $(H,h)$ and $(G,g)$ in \cref{eq:mpc2}.
But the \ac{MPC} can also implement plug-and-play support, expanding the matrices based on the number of attached inputs and outputs, and can replace functions in \cref{eq:mpc} with arbitrary resource heavy linear or non-linear alternatives. 
Offloaded controllers can also try different solvers, and parallel solutions, potentially utilizing distributed optimizations, as made available by the availability of cloud.
The potential for variation is large and \replace{real-time support can be problematic as it relies on known properties and and pessimistic assumptions, such as the worst case execution time of a task}{if used, it can be a problem to maintain sufficient real-time support}.

\subsection{\acl{R-CCS}} \label{sec:solution}
\begin{figure}[t]
		\centering
  		\insertfigure[tikz]{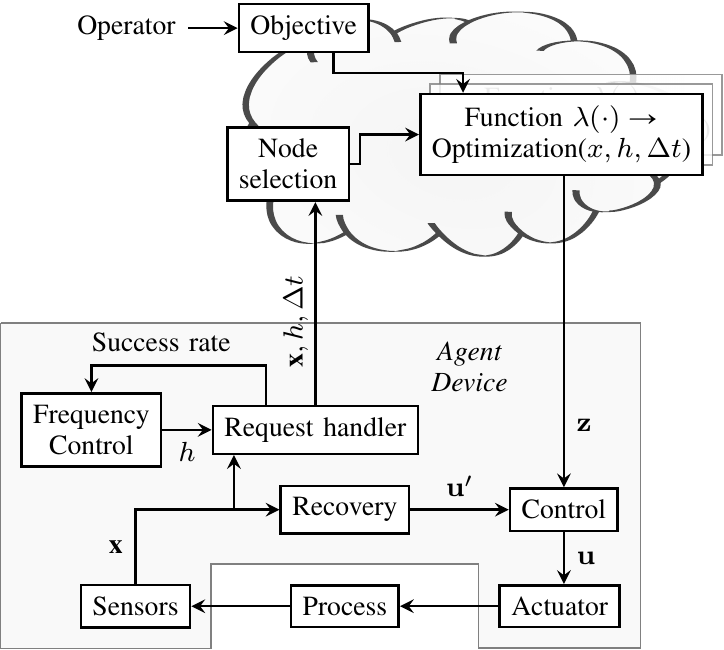}
      	\caption{The implemented \ac{CCS}. Nominal control is provided by an \ac{MPC} implemented by offloading to the cloud. Frequency control is used to modify the sampling period ($h$) of this controller, mitigating execution time dynamics caused by the controller itself and external factors. The full predictions of the \ac{MPC} are passed back to the device and used to handle latency from network and execution time. A recovery mode is provided for when the latency is too large for reliable control.}
  		\label{fig:cloud-control-solution}
  	\end{figure}

    The challenges in \cref{sec:preliminaries} are addressed by our proposed \ac{R-CCS}, a model of which is shown in \cref{fig:cloud-control-solution}.
    \cref{fig:cloud-control-solution} shows how an advanced, high performance controller is deployed for offloading in the cloud.
   	A local feedback system, the \emph{recovery controller}, ensures that timely control input is available at the plant, should the cloud controller be unreachable or fail to produce a timely result.
    Our goal is to achieve good performance and not send excessive requests to the cloud. 
    Below we detail the four remedies that constitute the \ac{R-CCS}.
  
%

    \subsubsection{Rate switching}
		Rate switching refers to the active change in the request rate of the cloud controller.
		When the rate changes, the \ac{R-CCS} also changes the operating frequency of the controller. 
		A reduced control frequency has the dual effect that \cref{eq:mpc} is given a longer deadline ($T$) while the optimization is also solved faster.
		This allows for the \ac{R-CCS} to simultaneously adjust its load on the network and in the cloud, while adjusting to a frequency that provides timely responses. 

    \subsubsection{Latency mitigation} \label{sec:openloop}
    	The remote controller is a predictive controller.
    	In a nominal \ac{MPC}, a single control action ($u_k(0)$) is applied from the resulting sequence $[u_k(0),\dots,u_k(N-1)]$.
    	The procedure is repeated with a new optimization in every sampling period.
    	In this mode, the cloud function only has to respond with this single control action.
    	In the \ac{R-CCS}, state and control signal predictions are transferred in $\mathbf{z}$ to the plant,
    	allowing for strategies that use predictions to mitigate response latency.
    	Here we use the basic approach of applying the predicted control actions, referred to as open-loop control.
    	More advanced alternatives are not considered here as they do not substantially add to the discussion.
	
            
    \subsubsection{Synchronization} 
        To simplify implementation and limit the effects of switching, control events are synchronized. 
        This is accomplished by the time slotted system illustrated in \cref{fig:timing}\replace{, also showing the use of open loop actions and change of control frequency}{}. 
   		Rows in \cref{fig:timing} show responses with indexed slots representing control actions. 
   		Gray slots show the time slots that are used. 
   		The selected control actions are shown on the horizontally axis as $u^i_k$ where $k$ is the sample index and $i$ indexes the predictions for this sample ($u^i_k$ is also written as $u_k(i)$).
		\begin{figure}[t]
			\vspace{-0.5em}
			\edef\figureoptions{:showresponsetime:nodeadtime:nolag:hqright}        	
			\insertfigure[tikz]{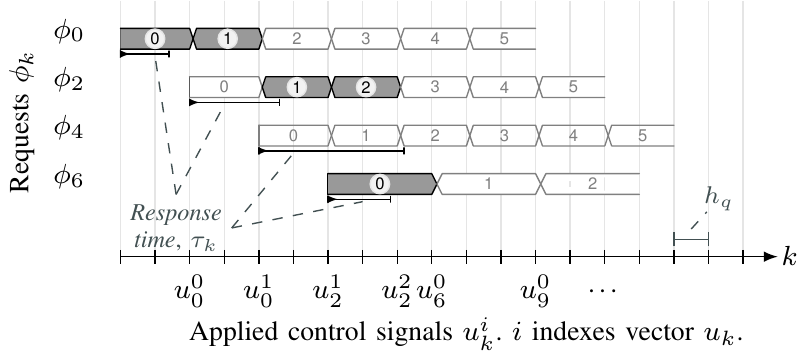}
			\vspace{-0.5em}
			\captionandlabel{Time slotted selection of controller result. Shaded slots show used control signals.\replace{ On the x-axis selected control signals are written as $u^i_k$, with $k$ the sampling time and $i$ the prediction index.} At $k=6$ the control frequency changes and the response to $\phi_4$ is discarded in favor of the new frequency.}{fig:timing}
			\vspace{-1em}
		\end{figure}
   		 
        The client is capable of sampling the plant and adjusting the control signal at the \textit{base rate} $\gls{basefreq}=1/\gls{stepsize}$.
        The \ac{MPC} period, $h_d$, is adjusted as multiples of the time step $h_q$, making the instantaneous request rate $\gls{reqfreq} = 1/(i \gls{stepsize})$ for some $i \in \mathbb{Z}^+$.
        The continuous time, $t$, is $t = kh_q$. 
    	A dead-time, $d = 1/\gls{reqfreq}$, is introduced so that the control signal $u^0_k$ is calculated for time $k\gls{stepsize}+1/\gls{reqfreq} =  (k + i)\gls{stepsize}$.
  		Ideal conditions are assumed for the agent, i.e., the reading of sensor values and the application of control signals can be perfectly timed.

		

           
\ifshowremoved
\markstartremoved
    \subsubsection{Synchronization and rate switching}
        The cost function is defined as
        \begin{equation}
        	\begin{aligned}
        		J(x_j, u_j, s_j, t) &= x_j^T \gls{gammax}(t) x_j  + u_j^T \gls{gammau}(t) u_j + \gls{gammas}(t) \gls{softc}_j
        		\label{eq:costfunction}
        	\end{aligned}
        \end{equation}	
        where the cost matrices $\Gamma_x$, $\Gamma_u$ and $\Gamma_s$ respectively penalize the plant state, the control inputs, and the slack variables.
        These matrices must be sampled for different frequencies.
        Similarly, $f$ is a linear state space model
        \begin{equation}
        	f(x,u,t) = \gls{A}(t) x + \gls{B}(t) u
        \end{equation}
        where the matrices $A$ and $B$ must be sampled differently at different times.
        The work in~\cite{skarincdc2020} provides a background on sampling and switching between different realizations of a system.

        \reviewwilliam{To me this section is both the challenges and the solution. So it think it should be split up. Do you agree?}
        As detailed in \cref{sec:targeted-system}, the system is consists of a client at the plant and a set of cloud-hosted \ac{MPC}, exposed as web-services, referred to as cloud deployments.
        In this scheme, the sensors and actuators are read and manipulated from the same client so determining the round-trip time between the client and a cloud deployment $\tau_k$ is trivial.
        The times $\tau^{sc}_k$, and $\tau^{ca}_k$ are combined to form a flight time delay $\tau^{fl}_k = \gls{tau}_k-\tau^c_k$.
        The processing time $\tau^c_k$ is measured by the application.
        Placing our concerns with $\gls{tfl}$ and $\tau^c$, there is no need for clock synchronization between client and cloud, but results extend directly to when sensors, controller and actuators are not co-located, with the orderly assumption of synchronized time.
        
        {\color{red} NOTE: Copied to B $|$ However, to simplify implementation and analysis, and limit the effects of switching, it is also necessary to synchronize all control events. 
        This is accomplished by working in multiples of a base frequency.
        The only limit to the rate of this frequency is that the client must be able to wake up and handle events.}

        \cref{fig:timing} shows this timing in a sequence of optimizations and helps in defining the two time frames, \textit{dead-time}, \gls{deadtime}, and \textit{control lag}, \gls{lag}, and highlight two limitations of the study.
        Each row in the figure shows one request, $r^k$, initiated from the sample collected at time $k$.
        Columns illustrate the output sequences from the \ac{MPC} calculations, with circled, black indexes showing used control signals while grayed slots are not used.
        Labels on the horizontal axis shows the applied control signal at tick $k$. 
        $r^0$ to $r_9$ shows a hypothetical sequence of events resulting in a new frequency, while $r'$ and $r''$ show two possible alternative outcomes.
        The control lag, $L(r^x_y)$, is the time from reading sensory input for request $x$ to the time of applying its calculated output at index $y$.
        The dead-time $D(r^x)$ is the time from reading sensory input to the activation time of the first control input $r^x_0$.
        The dead-time in the first six rows is $1/f_c(x)$. The seventh row, showing $r''$ exemplifies adding additional dead-time prediction to the request.
        This is a complement and alternative to changing frequency, but we limit the study to show a \ac{CCS} working also without this feature.
        
        A second limitation is seen when applying results from $r^2$.
        This response arrives after $k=4$ so its first input, $r^2_0$ is marked as obsolete.
        The input $r^0_1$ is applied in open loop because the response to $r^2$ has not arrived.
        When $r^2$ has provided a response its results are used because it has the shortest lag $L(r^2_1) < L(r^0_2)$, i.e., the most recent plant state as input to its calculations.
        When the open-loop prediction $r^0_1$ and feedback value $r^2_0$ are dissimilar, the error correction in $r^2$ can be of benefit but could possibly also cause further disturbance.
        To shorten the discussion and keep the implementation simple, we let the experimental results show if this is a major issue.

    	The conceptual setup of a \ac{CCS} is shown in \cref{fig:cloud-control}.
    	The \emph{client} is responsible for the function of a dynamic system referred to as the \emph{plant}.
    	To control the plant, the client implements an immutable \emph{recovery controller} and a \emph{primary controller} which is executed on a cloud platform.
    	The recovery controller executes on board the client at a fixed rate of $1/h$ Hz, with real-time guarantees, and timed with the base frequency.
    	This on-board controller is only for recovery, it is unable to achieve the performance goals of the primary controller.
    	
    	\reviewwilliam{The below paragraph should be in the targeted-system/model section}
        The execution on the client is near deterministic, the delay from sensor to actuator passing through the recovery controller ($\tau_{s \rightarrow a|u_l}$) is negligible and assumed zero. 
        However, the primary controller is affected by another dynamic system: the cloud.
    	The sensor to actuator delay of the primary controller, $\tau_{s \rightarrow a|u^*}$, referred to as the control lag, is not constant nor deterministic. \reviewwilliam{constant vs. deterministic?}
    	In addition, there is a non negligible cost of executing remote requests.
    	It is therefor of interest to the client that the offloading executes with the least possible uncertainty and at that the request rate is kept low. \reviewwilliam{That every request has value, i.e. loss is bad -> money down the drain.}
    	
\markendremoved
\fi

    \subsubsection{Frequency control} \label{sec:freqadapt}
        

		Frequency control is the process that determines the current rate of control actions. 
        This rate is aligned with the request rate.
        The solution continuously adapts the control period $h_d$ of the \ac{MPC} to achieve a tolerable percentage of request failures, represented as a miss ratio, $l$.
        The cost function in \cref{eq:mpc} becomes
        \begin{equation}
        	J(z, h_d(t)) = \sum_{j=0}^{N-1} J_s(z_j, h_d(t)) + J_f(z_N, h_d(t)),
        \end{equation} 
         where $J_t$ is a stage cost and $J_f$ a terminal cost.
         $h_d(t)$ is parameterized with $t$ to show that it can change between invocations but $t$ is constant in the optimization.
         The plant model, the initial state in the optimization, and the controller horizon is also affected to form
		  \begin{subequations}\label{eq:freqadaptivempc}
		  	\begin{align}
		  		x_{j+1} &&=&& f(x_j,u_j, h_d(t)),\\
		  		x_0 &&=&& \hat{x}(x(t),\mathbf{u}(t),x_s,h_d(t)),\\
		  		N &&=&& \ceil{N_c/h_d(t)}
		  	\end{align}
		  \end{subequations}
		  where $\hat{x}$ is a state estimate, $\mathbf{u}$ are control signal predictions, and $x_s$ is the setpoint.
		  $N_c$ is the controller horizon expressed in continuous time. 
		  
		The deadline miss ratio $l$ is updated with one miss or one hit every time step, $k$, of the base rate $h_q$.
		The hit or miss at time $k$ is determined by whether the most recently selected response has passed its sampling period deadline.
		In this way, the miss ratio update of a request is spread over $h / h_q$ base sampling periods and a response arriving at time $k$ with weight $w = h/h_q$ will have fully added its contribution to the miss ration at time $k+w$. 
		As such, it is a measure of delay.

        Building on works such as \cite{ploplys2004closed,xia2007flexible,bjorkbom2010wireless} \ac{PID} control is used to modify the effective sampling rate of the \ac{MPC}.
        First, a loss measure is defined as a smoothed version of the miss ratio using an exponential moving average filter
        \begin{equation}
        	\gls{reqloss}(t)  = \gls{emaalpha} \gls{miss}(t) + (1-\alpha)\rho(t-h_f).\label{eq:ema}
        \end{equation}
        where $l$ is the miss ratio over the time interval $h_f$ and $h_f$ is the freely selected sampling period of the controlling \ac{PID}. 
       	An acceptable loss, $\rho_r$, is defined to form the input, $e$, to the \ac{PID}
		\begin{equation}
			\gls{err}(t)  = \gls{losssp} - \rho(t).\label{eq:pidinput}
		\end{equation}       
		The \ac{PID} modifies the sampling period as
    	\begin{equation}
			\delta \gls{hc}(t)  = - \gls{pidk} \left( \dot{e}(t) + e(t) / \gls{pidti} + 	\gls{pidtd} \ddot{e}(t) \right),
    		\label{eq:pidcont}
    	\end{equation}    	
    	using standard dot notation for derivatives. \gls{pidk} is gain, $T_i$ the integration time constant, and $T_d$ the derivative time constant.
    	The \ac{MPC} period $h_d$ is determined from $h_c$ and $h_q$.

            	
    \subsubsection{Recovery} \label{sec:gohome}
        The recovery controller in \cref{fig:cloud-control-solution} provides the control action $u_l$ based on a locally calculated control problem.
        This provides a fallback mode when the primary controller is unavailable.
        The local controller can replace a malfunctioning controller with a simpler, lower performance, alternative, as in \cite{seto1998dynamic} and \cite{skarin2020cloudperformance}, but here we use a conservative, \textit{go-back-home} approach, where the recovery controller's objective is to bring the system to a safe initial configuration and stay there.
        This makes the study more expressive.
        A maximum control latency ($\gls{maxlag} \in \mathbb{R}^+$ ) determines when to apply recovery control.
        Combining predictions (\cref{sec:openloop}) with simple arbitration, the control block in \cref{fig:cloud-control-solution} selects a control action at time $t$ as
        \begin{equation}
        	u(t) = \left\{ \begin{matrix} 
        		u_{k|f_c(t)}(i|t), & \tau_t \le \gls{maxlag}, &i=[0,N-1] \\
        		u_l, & \tau_t > \gls{maxlag} &
        	\end{matrix} \right.
        	\label{eq:sigmathreshold}
        \end{equation}
    	where $f_c(t)$ is the active control frequency at time $t$ and $\tau_t$ is the delay of the remote control signal.
        We have been pragmatic and chosen a recovery controller that slowly progresses towards the origin.
        The effect of this will become evident in the experiments.


\subsection{Simulation}\label{sec:simsetup}
To study the effect of the remedies in \cref{sec:solution} and the hypothesis that rate control improves the performance of the controller while the system remains stable, simulation is used.
Input to the simulation experiments is borrowed from~\cite{skarin2020cloudperformance}.

\subsubsection{Processing time and network delay distributions}
SciPy~\cite{2020SciPy-NMeth} was used to turn data from~\cite{skarin2020cloudperformance} into the distributions in \cref{fig:distributions} using maximum-likelihood estimation.
The identified distributions and parameters are shown in \cref{tbl:dists}.
There are three different distributions: log-normal, generalized logistic, and double gamma.
The graph to the left in \Cref{fig:distributions} shows two distributions identified for task processing time when subjecting our RDC cloud (see \cref{sec:experiments}) to a large variety of \ac{MPC} optimizations.
These modes appeared randomly in the data.
In simulation, a processing time $\tau_c = i N X$ can be determined from the controller horizon $N$, the number of iterations in the optimization and drawing $X$ from one of these distributions. 
The graph on the right of \Cref{fig:distributions} shows task flight times from three different measurements.
Here, $s_3$ and $s_5$ are from two Kubeless deployments, one in the municipal RDC cloud and one in a distant public cloud (AWS in \cref{sec:experiments}). 
The flight times towards the distant data center ($s_5$) has two distinct modes of unknown origin but is likely to result from alternating worker nodes.
The third scenario, $s_4$, is from a cluster of machines implementing the optimization service using Flask and HAProxy.
The wide distribution of $s_4$ is from a high request load and represents a scenario of network congestion.
\newcommand\pflt[1]{\ensuremath{\pgfmathprintnumber[.cd,fixed relative,precision=3]{#1}}}%
\newcommand\pint[1]{\ensuremath{\pgfmathprintnumber[.cd,fixed,precision=0]{#1}}}%
\newcommand\pflsci[1]{\ensuremath{\pgfmathprintnumber[.cd,sci,precision=2]{#1}}}%
\newcommand\pflthp[1]{\ensuremath{\pgfmathprintnumber[.cd,fixed,precision=6]{#1}}}%
\begin{figure}[t]
	\centering
	\insertfigure[tikz]{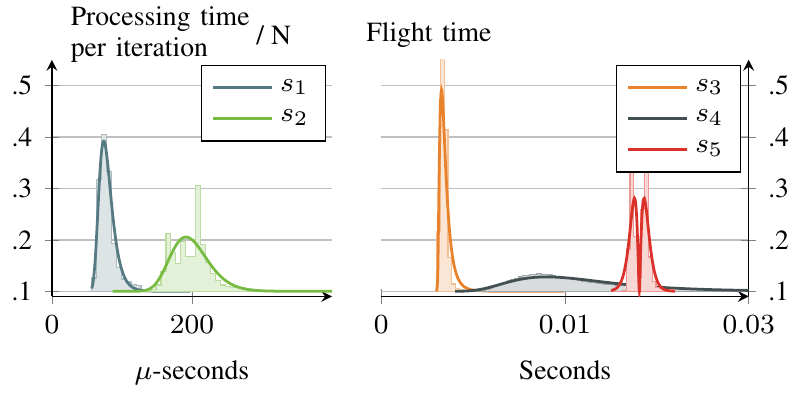}
	\vspace{-0.5em}
	\captionandlabel{Maximum likelihood fitted distributions used in the simulations.}{fig:distributions}
\vspace{-1em}
\end{figure}

\newcommand\psci[1]{\pgfmathparse{#1}\pgfmathprintnumber[.cd]{\pgfmathresult}}
\newcommand\p[1]{\pgfmathparse{#1}\pgfmathprintnumber[.cd,fixed,skip 0.]{\pgfmathresult}}
These distributions are combined with the transition probabilities in \cref{tbl:markov} to form Markov processes for processing time and flight time in two scenarios. 
The first scenario is chosen to have low transition probabilities so that states $s_1$ and $s_2$ represent an under- and over-utilized state in which the system is likely to remain for some time.
The second scenario is a representation of what was observed in the measurements.
Here transitions are likely in every iteration, creating a scenario where tasks end up randomly on different machines that execute at different speeds.
Task are handled twice as often by the slower machines represented by $s_2$.
\begin{table}[t]
	\centering
	\caption{Parameters for the distributions in \cref{fig:distributions}\per{Kan vi anse det här standard och inte ange någon eq eller ref?}}
	\label{tbl:dists}	
	\begin{tabular}{lllll}\toprule
		\multicolumn{2}{l}{State and Distribution} & \multicolumn{2}{l}{Parameters} & Offset\\\midrule
		$s_1$ & gen. logistic & $c=\pint{946.0079364124904}$ & $s=\pflsci{8.909250664719042e-06}$ & $\pflsci{1.2980603116739847e-05}$\\		
		$s_2$ & lognormal & $\sigma=\pflt{0.1928129897332288}$ & $\mu=\pflt{-8.864407134946601}$ & $\pflsci{5.5267691650919316e-05}$\\
		$s_3$ & lognormal & $\sigma=\pflt{0.63186}$ & $\mu=\pflt{-7.143477612503207}$ & $\pflsci{0.00602}$\\
		$s_4$ & lognormal  & $\sigma=\pflt{0.43087}$ & $\mu=\pflt{-4.2013063592469}$ & $\pflsci{0.0055629}$\\
		$s_5$ & double gamma & $a=\pflt{2.52199}$ & $s=\pflsci{0.00034}$ & $\pflt{0.02809}$\\ \bottomrule
	\end{tabular}
\end{table}
\begin{table}[t]
    \vspace{-1em}
	\centering
	\captionandlabel{Transition probabilities $p_{ij}$ from state $s_i$ to state $s_j$. Transitions are evaluated in every step of the simulation.}{tbl:markov}	
	\begin{tabular}{lcc|lc}\toprule
		& Scenario 1 & Scenario 2 & & Scenario 1 \& 2\\\midrule
		$p_{12}$ & \psci{0.001} & \p{0.75} & $p_{34}$ & \psci{0.00009}\\
		$p_{21}$ & \psci{0.001} & \p{0.25} &$p_{35}$ &\psci{0.00001}\\
		&&& $p_{43}$ & \psci{0.00025}\\
		&&& $p_{45}$ & \psci{0.00025}\\
		&&& $p_{53}$ & \psci{0.00035}\\
		&&& $p_{54}$ & \psci{0.00015}\\\bottomrule
	\end{tabular}
\vspace{-1.5em}
\end{table}

\subsubsection{Disturbances}
The simulations subject the system under control to periodic setpoint changes and introduces significant pulse disturbances that the controller also has to counteract.
The combination of disturbances, and small model errors, can force the system to break constraints and the optimization become infeasible.
To handle this, the standard application of soft constraints is used.
Disturbances enter the system as
\begin{equation}
	x(k+1) = Ax(k) + Bu(k) + \begin{bmatrix} 0 &  w(k) 	& 0 \end{bmatrix}^T.
\end{equation}
Here, $A$ and $B$ are the state space matrices for the base frequency, discretized using $h_q$ (\cref{fig:timing}).
The sequence of disturbances is, 
\begin{equation}
	\mathbf{w} = \begin{bmatrix} 0 & \dots & w_0 & 0 & \dots & w_1 & 0 & \dots  \end{bmatrix}, 
\end{equation}
where the values of $w_i$ are drawn from a normal distribution, $w_i = \mathcal{N}(\mu, \sigma^2)$, with $\mu = 0$, $\sigma^2 = 0.09$.
The position in $\mathbf{w}$, i.e., the time step $k$, is obtained as $k(w_i) = \ceil{t(w_i)/h_q}$, where the time steps $t(w_i)-t(w_{i-1})$, are drawn using another normal distribution with $\mu = 2$, $\sigma^2 = 0.25$.
Using this method, the amplitude and the timing of the disturbance varies. On average, disturbances occur with a distance of two second.

\section{Related work} \label{sec:relatedwork}

    Sampling and delays are old problems in control, dating back to the introduction of implementations using digital computers~\cite{dorf1962adaptive}.
    The effects of sampling and delays have been studied extensively, often in the design of systems with analytical solutions~\cite{cervin2003does,naghshtabrizi2006anticipative}, and scheduling in embedded and multiple processor systems~\cite{00777447,Pazzaglia-DATE2021}, but also for optimizing controllers~\cite{esen2015control}.
    Optimal control over networks with unknown delays was studied in 2000~\cite{8082073} and it is well known that predictive control can be used to mitigate unknown delay~\cite{0142331205tm133oa,04148044,miklovivcova2011model,ploplys2004closed}.
    Common to these examples is that the controller acts periodically at a set frequency, that latency is bounded and that controller performance is affected by assumptions on the delay distribution.
    A useful alternative is to redefine the systems to be event based~\cite{zou2016multiratempc,arzen1999pid}.
    Event based systems can reduce the communication load caused by the controller and naturally adapt the frequency of control.
    This has also been a motivation for predictive control with variable horizon in~\cite{ploplys2004closed}.
    However, once the control event triggers there can be strict requirements on the control actions and most examples focus only on network delay, they do not consider the execution platform as a shared resource.
    
    Because of significant uncertainty and variability~\cite{kim2005network,0142331205tm133oa,ploplys2004closed,xia2007flexible}, delay mitigation and adaptation are of particular interest to control over public and wireless networks.
    Recent works target modern concepts such as \ac{IIoT} and Fog computing~\cite{07879156,inaltekin2018virtualized}. 
    These works seldom include optimizing controllers and do not cover the possible effects, detrimental to the outcome of the system, caused by admission times and execution time delay~\cite{FINDEISEN2004427,Cortes2012,skarin2020cloudperformance}. 
    Although there are recent works that propose and study real-time closed loop control over the cloud~\cite{2FXjsvTrCQM3UqVVhvy3UB,4FwGZKjjHNC6ILk8Hg5bGE,Heiling2015}, few works provide reproducible insights with real-world data and details on the control system.
    This work fills that gap by introducing a complete control architecture, performance evaluation through simulation based on real data, and verification in the targeted cloud environment.
  
    This paper deploys software on what can be viewed as contemporary vanilla cloud environments.
  	The implementation specifically targets the use of function services, i.e. \ac{FaaS}.
  	It is therefore of particular interest that there is ongoing work to make these platforms more reliable and predictable.  
	There is ongoing research seeking to implement real-time support in cloud environments.
    \cite{Gupta2018,Perianayagam2022,Cucinotta2021} provide examples of implemented real-time storage and scheduling, while \cite{Szalay2021} propose a real-time scheduler to implement RT-FaaS.
    While these developments exist, they require careful planning and there is still a problem of \william{@PER!: verb?} end-to-end latency is service chains.
    In \cite{4FwGZKjjHNC6ILk8Hg5bGE} the authors implement control of a quad copter and illuminate the issue of latency when using service chains for stateful serverless computing.
    Cloudburst and Boki~\cite{Sreekanti2020,Jia2021} are two examples that aim to mitigate these issues.
    Both use local caches to reduce delay but they implement different approaches to consistency, elasticity, and resiliency. 
    Cloudburst is implemented on top of a scalable \ac{KVS} while Boki implements ordered updates to a shared log book.
    These examples can improve the average experienced delays and reduce latency tails but they do not provide guarantees and can result in unusually long delays in certain conditions.
\section{Experiments and results}\label{sec:results}
\newcommand\queuemark{$\dagger$}

        In this section, we present a set of experiments with the objectives to, 
\begin{enumerate*}
	\item \emph{validate and parameterize the frequency adaptation controller},
	\item demonstrate the \emph{individual and combined effectiveness of the methods} presented in \cref{sec:solution} to meet the challenges of resiliency, control performance and costs, and
	\item demonstrate that the \ac{R-CCS} is also able to
	\begin{enumerate*}
		\item cope with \emph{performance transients in the infrastructure},
		\item prevent \emph{resource starvation}, and
		\item successfully \emph{transition between heterogeneous cloud platforms}.
	\end{enumerate*}
\end{enumerate*}
The experiments demonstrate the feasibility of the proposed \ac{R-CCS} using realistic scenarios but do not to prove stability or determine absolute performance bounds, as this is very application dependent.

\subsection{Experiment setup} \label{sec:experiments}    

    The methods presented in \cref{sec:solution} are implemented as a \ac{R-CCS} in a process/client-side agent.
    For reproducibility, the code for the proposed methods and the data from the experiments will be made publicly available upon publication.
    This section proceeds with detailing the experiment setup followed by the setup of the experiments listed above and a summary of the parameters used in the experiments.
    
    The experiments were conducted using a set of heterogeneous cloud deployments and a \plant{}, simulated in real-time and controlled by a set of remote controllers deployed in a set of clouds.
    The metric in these experiments is the relative set-point tracking performance of the different configurations.
    The remedies from \cref{sec:solution} and the gain from using rate adaptation are first evaluated through simulation and then validated in cloud native software.
    
    To exercise the proposed \ac{R-CCS}, all experiments are subjected to noise and  disturbances throughout the duration of each experiment.
    Further, Gaussian noise, $\mu=0, \sigma=3$, is added to the control signal $u$.
    Additionally, the set-point is altered between $(-0.5, 0.5)$ every $\SI{10}{s}$.
    To be able to compare results and reproduce the results, noise is generated with a fixed seed, across all experiments.
    Finally, unless otherwise stated, all experiments run for 120 seconds. 

    \subsubsection{Experiment environment}
        The experiment environment consists of a process co-located with an agent that implements an \ac{R-CCS} and a controller implementation deployed to four cloud platforms.
        \begin{description}
        \item[Process]
            The \plant{} is the  \emph{Ball and Beam}~\cite{bnb} process, with the intuitive objective of moving and keeping a ball at a set point on a beam, without the ball falling off. The control input is the angular velocity of the beam. The outputs are the position of the ball and the angle of the beam.
            A plant-side agent, implemented in Python, handles logic and messaging.
        
        \item[The cloud-based controller and its implementation]\label{sub_sub_sec:controller}
            The remote service for the controller from \cref{sec:controller} is implemented in Python, using CVXOPT as the optimization framework.
            The implementation is accessed over a persistent HTTP connection. 
            A POST request containing the state of the plant should yield a response containing $u_k$.
            
        \item[Cloud platforms]\label{sub_sec:platforms}
            We deploy the remote controller implementation from \cref{sec:solution} to four different cloud infrastructures, heterogeneous in capacity, hosting platform, and network proximity.
            The infrastructures are:
            \begin{description}
                \item[K8S] a Kubernetes \cite{k8s} cluster of seven bare-metal nodes 
                adjacent to the process, our baseline deployment. The controller service is exposed using an nginx ingress controller. It has a measured median and \nth{95} quantile \acp{RTT} of \SI{11.71}{ms} and \SI{12.82}{ms}, respectively.
                \item[RDC] an Openstack-based research-focused \ac{DC} located 1.3 km (0.6 miles) from the process. The controller service is hosted on four \acp{VM}. Ingress traffic is routed through an HAProxy instance. Measured median and \nth{95} quantile \acp{RTT} are \SI{24.24}{ms} and \SI{26.46}{ms}, respectively.
                \item[Central \& North] are AWS Lambda functions with no throttling and with 1024 MB and 128 MB memory, respectively. They are hosted in \fraawsloc{} (\fraawsregion{}) and \stoawsloc{} (\stoawsregion{}),  respectively. The measured median and \nth{95} quantile \acp{RTT} are \SI{37.55}{ms} and \SI{52.62}{ms} for \fraawsregion{}, and \SI{180.06}{ms} and \SI{218.57}{ms} for \stoawsregion{}. The controllers  are exposed using AWS API Gateway.
            \end{description}
            
 		\end{description}
    \subsubsection{Experiments and metrics}
		\quad\\[1mm]
        \textbf{Validation of frequency adaptation strategy:} \label{sub_sec:freq}
            We desire the frequency adaptation strategy (presented in \cref{sec:freqadapt}) to find and settle at an appropriate control frequency (\gls{reqfreq}), given the experienced delay. A reactive adaptation with a smooth response is desired. This experiment validates the response of the frequency adaptation PID controller.
            The \ac{PID} in~\cite{xia2007flexible} is used as a baseline.
            Reconstructing its discrete form into \cref{eq:pidcont} yields the parameters $K = 3.5$, $T_i = 583$, $T_d = 214$.
			$h_c$ is obtained from the \ac{PID} and the effective request period, $h_d$, is  calculated as $h_d(t) = \left\lceil h_c(t)/\gls{stepsize} + 0.5 \right\rceil \gls{stepsize}$. The value is also limited by a sampling period range $h_{min} \le h_d(t) \le h_{max}$,
			When implemented, \cref{eq:pidcont} is approximated using first and second order forward approximations.            
            The following controllers are evaluated: \william{Why these particular $h_f$s?}
            \begin{enumerate*}\begin{enumerate*}
                \item PID, $h_f=0.1$
                \item PID, $h_f=0.5$
                \item PI, $h_f=0.5$,
            \end{enumerate*}\end{enumerate*}
        	where the last configuration has $T_d = 0$.
            To evaluate these frequency adaptation controllers, they are subject to a synthetic delay in a simulated environment.\\[1mm]
        \textbf{Effectiveness of proposed methods:} \label{sub_sec:effectiveness}
            The experiments aim to show the virtues of the individual methods and the value added when combined to an \ac{R-CCS}.
            The following configurations are evaluated: \textit{MPC} without any mitigating features\replace{, as in \cref{eq:mpc}}, \textit{a-MPC} as MPC with the recovery controller, \textit{oa-MPC} as MPC with recovery and open loop, and the \replace{proposed}{} system in \cref{fig:cloud-control-solution}, \textit{R-CCS}.\\[1mm]
            %
        \textbf{Infrastructure disruptions and transients:} \label{sub_sec:transient}
            These experiments are intended to exercise and validate the methods using a transient infrastructure, allowing for all system dynamics to interact constructively.
            The transient behavior is representative of events such as varying network load and consequences of infrastructure resource management policies.
            Building on the results from \cref{sub_sec:effectiveness}, here, we evaluate a \ac{R-CCS}'s ability to cope with disruptions in the infrastructure. 
            In these experiments we narrow the infrastructure down to the K8S cluster, which we have the most control over. 
            Further, the cluster is subject to a periodic delay using the Chaos Mesh \cite{chaos_mesh}.
            Specifically, the NetworkChaos module was used to subject one Kubernetes pod (a service instance) to a delay with a mean of $\SI{100}{ms}$, correlation of 25, and a jitter of $\SI{15}{ms}$, for $\SI{30}{s}$ every $\SI{1}{m}$. 
            To give the results contrast,  both a frequency adaptive controller and fixed frequency controller were used. 
            \replace{We expect that the frequency adaptive controller will in the ball better following the set-point, i.e. more accurate set-point tracking.}{}\\[1mm]
        \textbf{Resource starvation:} \label{sub_sec:starvation}
            The proposed \ac{R-CCS} is quality elastic which can potentially suppress resource starvation caused by contention between multiple cloud tenants (multi-tenancy), without fully compromising the performance of the tenants. 
            This aspect is relevant in terms of resource usage, but also in terms of resiliency, as it may take the cloud some time to respond to an increase in the workload. 
            In this set of experiments we are therefore interested in evaluating the impact of successively applying more \plant{}s (tenants) to a resource constrained instance of a remote controller (service). 
            Three \replace{identical }{}\plant{}s are used, admitted $\SI{20}{s}$ apart, all using a controller service deployed in K8S.
            At the start, the controller deployment is hosted on one pod, when all \plant{}s have started to send requests, the deployment is scaled to three pods, at $t=\SI{60}{s}$.
            In addition to the basic criteria of success, the success in this experiment is also determined by the ability to share resources, maintain stability, and track the set-point.\\[1mm]
        \textbf{Transition between cloud platforms:} \label{sub_sec:transition}
            Although the proposed \ac{R-CCS} is expected to cope with a transient cloud infrastructure, as the experiments in \cref{sub_sec:transient} aim to validate, any controller is has an upper \ac{RTT} limit, beyond which it is unstable. 
            If, at any point in time, that bound is breached, or another cloud can offer better performance in terms of control frequency, an \ac{R-CCS} shall be able to seamlessly switch to that cloud. 
            In this experiment, a new controller deployment is randomly selected among the cloud infrastructures detailed in \cref{sub_sec:platforms}, every $\SI{20}{s}$.
            The criteria of success is the \ac{R-CCS}'s ability to successfully transition between cloud deployments and timely adapt the control frequency of the process to that deployment.
            Note that the criteria for which cloud to employ, given a set of objectives and criteria, is beyond the scope of this paper.
        
            To mitigate delay and reduce the need for recovery control, the controllers in experiments 2-4 are configured with $\gls{sigma}=\SI{200}{ms}$ (see \cref{sec:openloop}).

    \subsection{Validation of frequency adaptation strategy} \label{sub_sec:freq}
\begin{figure}[tb]
			\insertfigure[tikz]{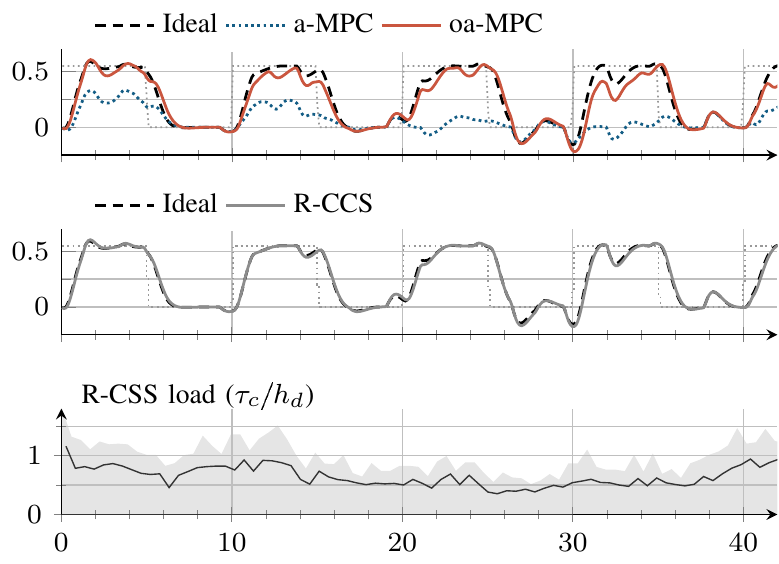}
            \caption{Time series outcome from a simulated environment, showing the benefits of frequency adaptation. 
            \replace{The top graph shows the position of the ball over time for two different controllers contrasted with the ideal outcome. The middle graph shows the potion of the ball using \ac{R-CCS} in contest to the ideal outcome. The bottom graph shows the average (solid line) and worst case (gray back drop) load as it would be experienced in the cloud.}{}
            }
        	\vspace{-0.5em}
			\label{fig:ccssim}
            \vspace{-.5em}
		\end{figure}

    	First, a set of simulations are used to establish the PID controller to use.
    	For brevity these simulations are excluded here but results in the selection of the PI configuration with $h_f = 0.5$ and a loss rate $\rho_r$ of 0.05.
    	Two things now need to be established in the extended simulations:
    	1) that the \ac{R-CCS} works reliably and 2) that the \ac{R-CCS} frequency adaptation provides a performance improvement.
    	A snapshot of the validation is shown in \cref{fig:ccssim}.
    	This example uses the distributions and Markov processes of scenario two in \cref{sec:simsetup}.
    	The top graph shows the result of three controllers. 
    	The first (dashed black) is an ideal result where the network and processing delay is zero, i.e., an \ac{MPC} that never misses its deadline. 
    	The second (dotted blue) is the result of an ordinary \ac{MPC} experiencing packet loss.
    	This result is far from the ideal.
    	The third (red) is a networked \ac{MPC} that uses its predictions to mitigate delay.
    	This works better, but is does not match the ideal, as evident by a visual inspection of \cref{fig:ccssim}.
    	The middle graph shows the result of the \ac{R-CCS} (gray).
    	The ideal is shown again in this graph and it is clear that the result is a major improvement in comparison to the top graph.
    	To see how the sample rate adaptation affects the \ac{R-CCS} the bottom graph shows the average (solid line) and worst case (gray back drop) load as it would be experienced in the cloud.
    	The versions in the top graph would show a flat line here.    	
		\newcommand\Ideal{Ideal}
		\newcommand\SDNO{SPSC}
		\newcommand\SDOL{SPOL}
		\newcommand\MDOL{NDOL}
		\newcommand\Adapt{Adapt}
		
\begin{figure}
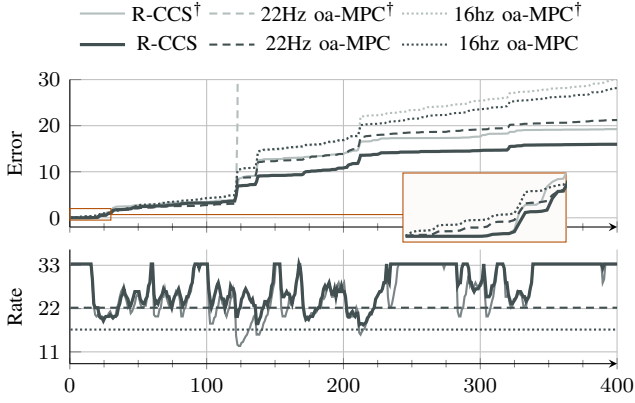

			\insertfigure[tikz]{clefigure}
			\caption{Comparing rate adaptation and fixed rate oa-MPC. Results in light gray, marked with $^\dagger$ in the legend, indicates a simulation with limited resources in terms of worker nodes. The dark gray results are ideal in that the system can handle any amount of requests in parallel.}			
			\label{fig:closedlooperr}
            \vspace{-1.5em}
		\end{figure}

		We now consider the potential for changing the oa-MPC's sampling period to improve its response.
		In the previous example the fixed rate controllers execute at \SI{33}{Hz}.
		We setup two alternatives at \SI{17}{Hz} and \SI{22}{Hz} to reduce the load.
		These values are based on the frequency observed for the \ac{R-CCS} in the simulations.
		We define an error measure, the \ac{CLRE}, using the controlled state $x_1$ in reference to the response of the ideal controller $\bar{x}_1$ as
		\begin{equation}\label{eq:cler}
			\text{CLRE}(t) = \int^t (x_1(s) - \bar{x}_1(s))^2 ds,
		\end{equation}
		Results for the two scenarios are shown in \cref{tbl:clre}.
		Here, examples are also shown represented with a $\dagger$.
		These are simulations that do not model an ideal cloud but instead assume a single worker node that can be overloaded.
		Including this alternative shows the strength of the \ac{R-CCS} in scenarios that might appear in relation to cloud and \ac{IoT}.
		From the table we see that the \ac{R-CCS} bests all static alternatives.
		\cref{fig:closedlooperr} also shows the build up of error over time and the applied sampling rate of the \ac{R-CCS}
 
		\begin{table}[t]
			\caption{Closed loop error response after 400 seconds simulation.}
			\label{tbl:clre}
			\begin{tabular}{lp{3em}p{3.4em}p{3em}p{3em}p{3em}p{3em}}\toprule
				& R-CCS & R-CCS\queuemark{} & 22 Hz&22 Hz\queuemark{} & 17 Hz& 17 Hz\queuemark{} \\\midrule
				Sc. 1& 15.96&19.25&21.22&307.87&28.14&30.00 \\
				Sc. 2& 20.19&21.06&24.77&25.10&30.15&30.15 \\
				\bottomrule\end{tabular}
            \vspace{-1.5em}
		\end{table}
	
    \subsection{Effectiveness of each proposed remedy} \label{sub_sec_result:effectiveness}
        We now turn to experiments on the cloud infrastructure. \cref{fig:sucesssivecomparison} shows the over-laid time-series of the outcome from the four configurations listed in \cref{sub_sec:effectiveness} when subjected to a disturbance, as defined in \cref{sec:experiments}.
        The top plot shows the most critical state of the plant, the position of the ball, subjected to a square-wave set-point for the controller to track.
        The plot below shows the dynamic frequency, $f_c$, for \ac{R-CCS} and the constant frequency.
        \per{Säga något om exekveringstid?} \reviewwilliam{@PER!:Load finns i figuren men det varken defineras eller analyseras i stycket.}
        
        
        \begin{figure}[t!]
        	\centering
            \insertfigure[tikz]{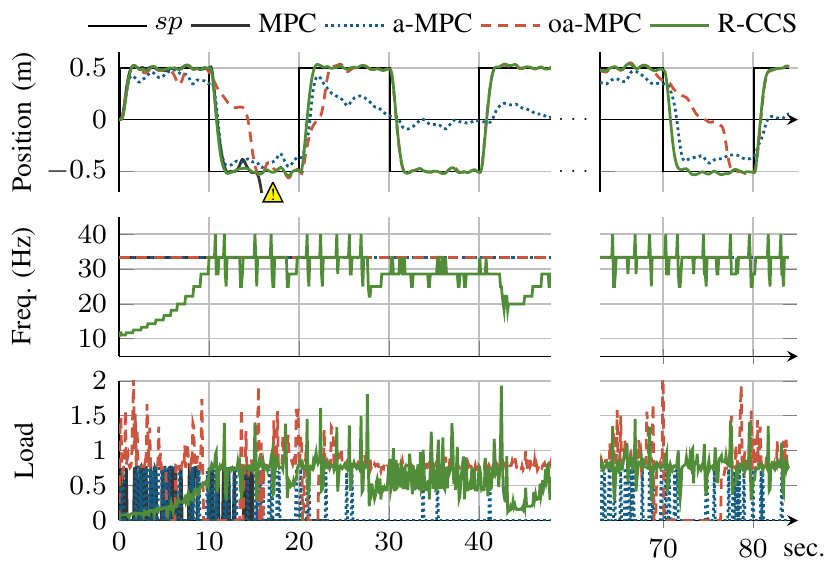}
        	\captionandlabel{Evaluation of the merits of individual remedies and combinations of remedies, including the proposed \ac{R-CCS}. $sp$ is the position set-point.}{fig:sucesssivecomparison}
        \end{figure}
        
        When controlled by the uncompensated MPC, the ball falls of the beam at the second set-point change due to successive losses, no ability to adapt, and no fall back on a local controller. This is emphasized in \cref{fig:sucesssivecomparison} by a circled exclamation mark.
        It shows that the predictive controller is insufficient to stabilize the plant under the conditions of the experiment, and in extension, the cloud.
        Adding the local recovery controller, yields the a-MPC configuration.
        This configuration is reliable, but far from satisfactory.        
        The recovery controller is conservative and often has to move the position towards the center, in competition with the MPC goal of reaching the set-point.
        As a consequence, the control loop remains stable but seldom manages to reach the set-point.                
        Introduction of an acceptable control latency, in configuration oa-MPC, implemented using the open loop sequence (\cref{sec:openloop}), provides an improved response.
        This configuration is able to keep the ball on the beam and track the set-point most of the time, but \cref{fig:sucesssivecomparison} also shows significant deviations from the set-point in the ranges $t \in (5,25)$ and $t \in (70,78)$.
        These problematic sequences are a result of prediction errors and losses beyond the maximum tolerable latency.
		Including the adaptive frequency controller in the configuration yields the \ac{R-CCS} which eliminates the aforementioned shortcomings and provides the desired outcome.
        It consistently shows a quick and smooth response to set-point changes and attenuates disturbances. \reviewwilliam{@PER!: Disturbance attenuation är inte definerat.}        
        In \cref{fig:sucesssivecomparison}'s bottom plot, the control frequency of the \ac{R-CCS} starts at \SI{10}{Hz} but successfully adapts within 10 seconds. 
		It settles at a median control frequency of $\SI{33}{Hz}$, equal to the other configurations.
		Later, from $t \approx 28$, response times increase and the \ac{R-CCS} changes to a lower control frequency while retaining the control performance.        

    \subsection{Synthetic disturbance to the infrastructure} \label{sub_sec_result:transient} \william{This is our weakest result, remove?}
        In \cref{sub_sec_result:effectiveness}, it was concluded that the proposed \ac{R-CCS} is able to adapt to the state of the infrastructure and therefore functionally outperforms the other configurations. 
        This set of experiments instead focuses on the proposed system's resilience to disruptions in the infrastructure and its effective use of resources.
        Note that, as specified in \Cref{sub_sec:transient}, the mean delay disturbance of $\SI{100}{ms}$ is below the maximum control latency of $\SI{210}{ms}$.
        Therefore neither configuration will intentionally be pushed beyond the point where it will not receive any responses. 
        However, when operating at $\SI{30}{Hz}$, the delay of $\SI{100}{ms}$ is greater than the sampling period.
        Thus, on average, an older control decision will be applied. 
       
        \cref{fig:freqdisturbance} shows the over-laid time-series of the outcome from using fixed (oa-MPC) and adaptive (\ac{R-CCS}) control frequencies, when subjecting the hosting cloud infrastructure to an intermittent disturbance, in the form of delay. How the delay changes is seen from the RTT in the second graph of the figure, and below that, the request rate, $f_c$, of the \ac{MPC} is shown.
        \begin{figure}[t!]
        	\centering    
            \insertfigure[tikz]{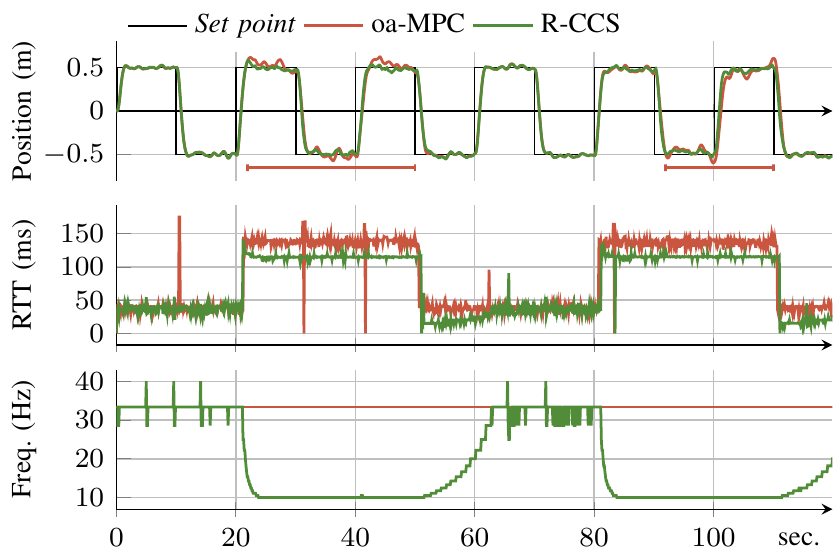}
        	\captionandlabel{Time series when applying delay disturbance to the cloud deployment, comparing fixed and adaptive frequency, as detailed in \cref{sub_sec_result:effectiveness}.}{fig:freqdisturbance}
        \end{figure}                
        The response is better, and the \ac{RTT} is noticeably lower, during the two periods of high delay when using \ac{R-CCS}.
        During those periods, the \ac{R-CCS} reduces the control frequency to match the experienced request \ac{RTT}.
        There is a double effect in that this allows more time for the closed loop response to arrive, while also lowering the system load, providing a shorter response time.
        The improved response of the \ac{R-CCS} is clearly visible in the marked ranges (in the color of the oa-MPC).
        For \ac{R-CCS}, over the course of the experiment, the accumulated error and execution time are $11\%$ and $51\%$ lower, respectively, compared to oa-MPC.
        A lower error and lower execution time translates to a higher yield per resource, lower resource usage, and potentially reduced costs.
        Being adaptive and occasionally operating at a lower frequency can be beneficial.

    \subsection{Resource sharing} \label{sub_sec_result:starvation}
        
\begin{figure}[t!]
        	\centering
            \insertfigure[tikz]{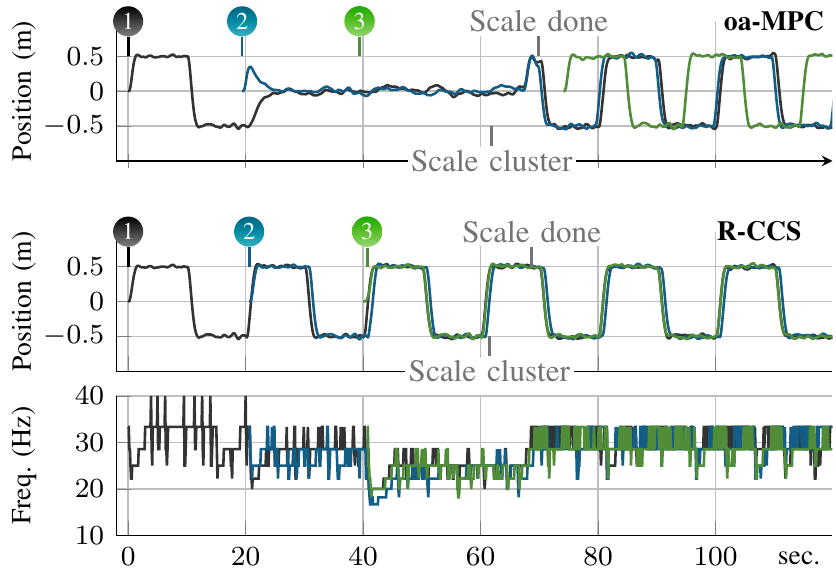}
        	\caption{Time series of outcome of resource starvation, comparing fixed and adaptive frequency, as detailed in \cref{sub_sec:starvation}.\reviewwilliam{What is loss?}\reviewwilliam{To reduce spacing between plots (group them): Split into two sub plots with a label each.}\reviewwilliam{Should probably have added more, to show that it is possible. :) What is the limit? 10?}}
        	\label{fig:compeeting}
        	\vspace{-1.25em}
        \end{figure}

        As alluded to in \cref{sub_sec_result:transient}, being able to adapt the control frequency dynamically can reduce the load on the hosting infrastructure. 
        That property is now further explored.
        \cref{fig:compeeting} shows responses for plants 1,2, and 3, successively admitted $\SI{20}{s}$ apart, all served by the same cloud service, hosted in K8S.
        The first and second chart show the outcome when using fixed control frequency (oa-MPC) and adaptive control frequency (\ac{R-CCS}), respectively.
        
        When using a fixed control frequency and only one \plant{} is using the cluster, ($0 \leq t < 20$), the cloud-deployment is able to accommodate the \ac{MPC} controllers at $\SI{33}{Hz}$. 
        At $t=20$, a second \plant{} is admitted, and at $t=40$, a third.
        There is a visible attempt by the second \plant{} to move towards the set-point but it quickly has to revert to the recovery controller which moves the position to zero.
        As all \plant{}s continue to attempt to request the cloud, it remains overloaded \reviewwilliam{Overloaded is not defined.} and none of the \plant{}s are able to track the set-point.
        At around $t=60$, the number of control service instances in the cluster \replace{(\cref{compeeting})}{} is manually scaled from one to three pods. 
        When scaling is done, all three \plant{}s can be accommodated and subsequently they leave the 'go-back-home'-mode and resume to successfully track the set-point.
        The delayed start of \plant{} three is due to the overloaded situation and that simulation of the plant is started with the first timely response.
        
        When employing adaptive control frequency, i.e., using the proposed \ac{R-CCS}, all three \plant{}s successfully track the set-point, with little or no error. 
        Independently but in tandem, the \plant{}s reduce their control frequency until they reach the tolerated level of loss, for each successive \plant{} admission.
        An unanticipated, positive, side effect is that the entry of the new \plant{} is not observable from the plant state, at least not from a visual examination of the position graph.
		Meanwhile, the entry of a new \plant{} very abruptly negatively affected the outcome when using oa-MPC. 
		The admittance of new \plant{}s is visible in the frequency graph.
		The external and persistent disturbance in \cref{fig:freqdisturbance} caused a fast frequency reduction and later a slow ascent back to the high frequency.
		Here we see how the \plant{}s quickly reduce frequency in tandem, then progress back to a useful frequency. While there is an overshoot to a slightly lower frequency the response is fast and the \plant{}s never fall down to the lowest frequency (of \SI{10}{Hz}).
   
        At $t=60$, the deployment again scales to three pods. 
        As shown in the bottom chart in \cref{fig:compeeting}, each \plant{} independently and opportunistically increase their control frequency. 
        The control frequency, after the deployment has scaled, is lower than with one pod and one \plant{}. 
        This is due to the increased load of three \plant{}s on the cluster as a whole.
        
        \begin{figure}[t!]
        	\centering    
            \insertfigure[tikz]{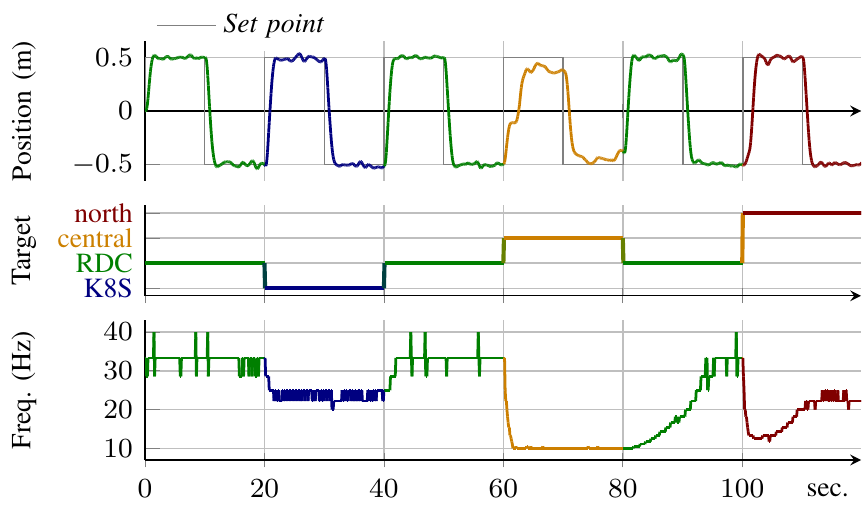}
        	\vspace{-1.5em}
        	\caption{Time series of outcome of one system transitioning between set of heterogeneous cloud platform.}        	
        	\label{fig:random_target}
        	\vspace{-1em}
        \end{figure}
        
    \subsection{Migrating between cloud platforms} \label{sub_sec_result:transition}
        Here, the proposed \ac{R-CCS}'s resilience through portability is demonstrated.
        One \plant{} randomly switches between the four cloud deployments detailed in \cref{sub_sec:platforms}.
        In \cref{fig:random_target} the upper chart shows the position of the ball, below, the target cloud, followed by the controller frequency, as time-series.

        As demonstrated in \cite{skarin2020cloudperformance}, the deployments have different performance properties. 
        Nevertheless, the \ac{R-CCS} is able to seamlessly switch between the deployments and expediently adapt to a control frequency appropriate to that deployment. 
        The deployment \emph{north} has the highest \ac{RTT} and the most varied execution time and is therefore the most challenging.
        The deployment \emph{K8S} has the lowest \ac{RTT} but does not achieve the highest frequency. It can be assumed that this is due to the longer computation time on that deployment. RDC is able to achieve the highest control frequency.
        At $t=60$, the control frequency is reduced due to an increase in \ac{RTT}. 
        Here we see how the client struggles to track the set-point when the recovery controller repeatedly forces the plant towards a safe state.
		After $t=80$ control frequency has adapted to the new \ac{RTT} and the system settles at a new performance level.

\section{Conclusions} \label{sec:conclusions}
	We have presented a resilient cloud controller design through an offloading architecture and frequency control.
	The output of a loss ratio based smoothing filter and PI control is used to manage a client request rate.
	The result implements constrained \ac{MPC} using cloud-native and \ac{FaaS} on private and public clouds.
	To ensure reliability, the client implements a \textit{go-back-home} mode for when the cloud is unresponsive. 
	Through experiments we show how the \acl{R-CCS} is able to mitigate delay and keep a process running when subjected varying delay, typical of a cloud-environment. 
    Further, it was also shown that \ac{R-CCS} allows for multiple controllers to co-adapt and share resources without interfering with each other, allowing a cluster to admit new controllers while scaling to meet the new demand. Also, it was shown that the process can survive abrupt migrations between cloud.
	These positive effects are achieved while at the same time keeping the number of unused, wasteful, requests low.
	Further work on the \ac{R-CCS} can introduce other means of altering the controller load on-line, built in delay mitigation, statistical methods for setting the frequency, and improved automation through ML.

\bibliographystyle{plain}
\bibliography{main}

\end{document}